\documentclass[useAMS,usenatbib]{mn2e}

\usepackage[dvips]{graphicx}
\usepackage{amssymb}
\usepackage[english]{babel}

\title[The Helium Abundance in Lower Main 
Sequence Stars]{The Helium Abundance and $\Delta Y / \Delta Z$ in Lower Main 
Sequence Stars}\author[Casagrande, Flynn, Portinari, Girardi, Jimenez]{Luca 
Casagrande$^{1}$, Chris Flynn$^{1}$, Laura Portinari$^{1}$, 
Leo Girardi$^{2}$, Raul Jimenez$^{3}$\\
$^1$ Department of Physics and Tuorla Observatory, University of Turku,
FIN-20014 Turku, Finland\\
$^2$ Osservatorio Astronomico di Padova, Vicolo dell'Osservatorio 5, 35122
Padova, Italy\\
$^3$ Department of Physics and Astronomy, University of Pennsylvania, PA 
19104, USA}
\begin{document}

\maketitle

\begin{abstract}

We use nearby K dwarf stars to measure the helium--to--metal enrichment ratio
$\Delta Y / \Delta Z$, a diagnostic of the chemical history of the Solar
Neighbourhood. Our sample of K dwarfs has homogeneously determined effective
temperatures, bolometric luminosities and metallicities, allowing us to fit
each star to the appropriate stellar isochrone and determine its
helium content indirectly. We use a newly computed set of Padova isochrones
which cover a wide range of helium and metal content.

Our theoretical isochrones have been checked against a congruous set of main
sequence binaries with accurately measured masses, to discuss and validate
their range of applicability. We find that the stellar masses deduced from the 
isochrones are usually in excellent agreement with empirical measurements. Good
agreement is also found with empirical mass--luminosity relations.

Despite fitting the masses of the stars very well, we find that anomalously low
helium content (lower than primordial helium) is required to fit the
luminosities and temperatures of the metal poor K dwarfs, while more
conventional values of the helium content are derived for the stars around
solar metallicity. 

We have investigated the effect of diffusion in stellar models and LTE 
assumption in deriving metallicities. Neither of these is able to resolve the 
low helium problem alone and only marginally if the cumulated effects are 
included, unless we assume a mixing-length which is strongly decreasing with 
metallicity. Further work in stellar models is urgently needed.

The helium--to--metal enrichment ratio is found to be $\Delta Y / \Delta Z = 
2.1 \pm 0.9$ around and above solar metallicity, consistent with previous 
studies, whereas open problems still remain at the lowest metallicities. 
Finally, we determine the helium content for a set of planetary host stars.

\end{abstract}


\section{Introduction}\label{intro}

K dwarfs are long-lived stars and can be regarded as snapshots of the stellar 
populations formed at different times over the history of our Galaxy, 
therefore constituting an optimal tool for any study dealing with its chemical 
evolution (e.g. Kotoneva et al. 2002).
K dwarfs share a similar metallicity
distribution with the G dwarfs, in which most stars have metallicities around
the solar value, a feature not expected in the simplest, closed-box models of
Galactic chemical evolution (i.e. the G-dwarf problem). In addition to the
abundance patterns of metals in a stellar population, the helium content $Y$
can also diagnose its chemical evolution, but this diagnostic has received less
attention because measuring the helium content of low mass stars can only be
done indirectly (e.g. Jimenez et al. 2003). Typically, one measures a 
differential production rate of the helium mass fraction $Y$ in the stellar 
population
relative to the metal mass fraction $Z$, i.e. $\Delta Y/\Delta Z$. For the
solar neighbourhood, Jimenez et al. (2003) determine $\Delta Y/\Delta Z \approx
2.1 \pm 0.4$ from K dwarfs, a value similar to that found by studying H~II
regions in both the Milky Way and external galaxies (e.g. Balser 2006).  Metals
mainly come from supernovae with high-mass progenitors, whereas helium is also
injected into the interstellar medium by mass-loss from intermediate and
low mass stars : $\Delta Y / \Delta Z$ can thus be computed from stellar
evolutionary theory for a given initial mass function (e.g. Chiosi \& Matteucci
1982; Maeder 1992, 1993; Chiappini, Renda \& Matteucci 2002). The ratio 
$\Delta Y / \Delta Z$ can also be used to infer the primordial helium 
abundance $Y_P$ by extrapolating to $Z=0$; the technique is usually applied 
to extragalactic H~II
regions -- the K dwarfs studied here provide an independent check on $Y_P$.
Cosmic Microwave Background measurements alone are not able to provide a tight
constraint in $Y_P$ (Trotta \& Hansen 2004), but WMAP3 data on the cosmic
baryon density when combined with Standard Big Bang Nucleosynthesis returns a
formally accurate value for $Y_P$ (see Section \ref{heres}).  Finally, age
determinations for both resolved and integrated stellar populations typically
assume a value for $\Delta Y/\Delta Z$, and accurately knowing the age of
galaxies can, in turn, help to determine the nature of dark energy (Jimenez \&
Loeb 2002).

Helium lines are not easily detectable in the spectra of low mass stars, with 
the notable exception of hot horizontal branch objects, whose atmospheres are, 
however,
affected by gravitational settling and radiative levitation, which strongly
alter the initial chemical stratification (e.g. Michaud, Vauclair \& Vauclair
1983; Moehler et al. 1999) and whose composition anyway would not reflect the
original helium abundance at their birth.  Therefore, assumptions have to be
made for the initial helium content of models of low-mass stars. Very often,
for the sake of simplicity, it
is supposed that the metallicity $Z$ and helium fraction $Y$ are related
through a constant ratio $\Delta Y/\Delta Z$. The latter is often determined
from the result of the solar calibration $(Y_{\odot}-Y_P)/Z_{\odot}$ and for
any other star of known metallicity, the helium abundance is scaled to the
solar one as $Y=Y_{\odot}+\frac{\Delta Y}{\Delta Z} \times (Z-Z_{\odot})$.
Most of the conclusions drawn when comparing theoretical isochrones with
binaries (e.g. Torres et al. 2002; Torres \& Ribas 2002; Lacy et al.  2005;
Torres et al. 2006; Boden, Torres \& Latham 2006; Henry et al. 2006) and field
stars (e.g. Allende Prieto \& Lambert 1999; Valenti \& Fischer 2005) thus
reflect this tacit assumption on the helium content.

Recent results, however suggest that the naive assumption that $\Delta Y /
\Delta Z$ varies linearly and with a universal law might not be correct.  For
example, the Hyades seem to be $Y$ deficient for their metallicity (Perryman et
al. 1998; Lebreton, Fernandes \& Lejeune 2001; Pinsonneault et al. 2003).  The
discovery that the Globular Cluster $\omega$~Cen has at least two different
components of the main sequence and multiple turnoffs (Lee et al. 1999; Pancino
et al. 2000, 2002; Ferraro et al. 2004; Bedin et al. 2004) can be explained
assuming stellar populations with very different helium abundances (Norris
2004) and thus very different $\Delta Y/\Delta Z$. Recently, compelling
evidence has been found for a helium spread among the main sequence in the
Globular Clusters NGC 2808 (D'Antona et al. 2005a) and the blue horizontal
branch stars in the Globular Cluster NGC 6441 (Caloi \& D'Antona 2007).  At the
same time, for a different sample of Galactic Globular Clusters Salaris et
al. (2004) found a very homogeneous value of $Y$ with practically no helium
abundance evolution over the entire metallicity range spanned by their study.
All this suggests that a patchy variation of $\Delta Y/ \Delta Z$ and complex
chemical evolution histories might be not so unusual in our Galaxy. Similar
indications also start to appear for extragalactic objects (Kaviraj et
al. 2007). There are various methods to infer the helium content in Globular
Clusters (e.g. Sandquist 2000) and all take advantage of the fact that in these
objects it is relatively easy to perform statistical analysis over large
stellar populations.

The determination of the helium content in nearby field stars, on the contrary,
is more challenging, because it is less straightforward to build a
statistically large and homogeneously selected sample. Accurate parallaxes are
needed and to avoid subtle reddening corrections only stars closer than $\sim
70$~pc must be used.
All studies to date have exploited the fact that the
broadening of the lower main sequence with metallicity effectively depends on
the helium content (see Section \ref{method}) so that its width can be used to
put constraints on $\Delta Y / \Delta Z$ (e.g. Faulkner 1967; Perrin et
al. 1977; Fernandes, Lebreton \& Baglin 1996; Pagel \& Portinari 1998; Jimenez
et al. 2003). In this work we take the same
strategy, comparing the positions of a large sample of field stars with
theoretical isochrones in the $M_{Bol} - T_{eff}$ plane. For the parameter
space covered by the isochrones, the number of stars used, the accuracy and the
homogeneity of the observational data (crucial when it comes to analyzing small
differential effects in the HR diagram), this is the most extensive and
stringent test on the helium content of lower main sequence stars undertaken to
date.

An analogous work was pioneered by Perrin et al. (1977) with 138 nearby FGK
stars but with much less accurate fundamental stellar parameters,
pre-\emph{Hipparcos} parallaxes and of course, older stellar models.  After
\emph{Hipparcos} parallaxes become available, the problem was re-addressed by
Lebreton et al. (1999) with a sample of 114 nearby FGK stars in the metallicity
range $-1.0 < \textrm{[Fe/H]} < 0.3$, of which only 33 have $T_{eff}$ and
$M_{bol}$ determined directly via the InfraRed Flux Method (hereafter IRFM) of
Alonso et al. (1995, 1996a). For the remaining stars, temperatures were either
recovered via spectroscopic methods or color indices and $M_{Bol}$ determined
from the bolometric corrections of Alonso (1995, 1996b).

We have recently carried out a detailed empirical determination of fundamental
stellar parameters via IRFM (Casagrande, Portinari \& Flynn 2006) with the
specific task of determining the helium content in dwarf stars by comparing
them to theoretical isochrones.  Our sample is similar in size to previous
studies, but i) it has a larger metallicity coverage ii) we have improved the
accuracy in the selection (see Section \ref{sample}), iii) we have carefully
and homogeneously determined the fundamental stellar parameters iv) focusing
particularly on stars (K dwarfs) where the helium content can be most directly
determined from the stellar structure models, since the effects of stellar
evolution play an insignificant role.

Stellar models are common ingredients in a variety of studies addressing
fundamental cosmological and astrophysical problems, from ages and evolution of
galaxies, to complex stellar populations, to exoplanets.  Nevertheless, our
incomplete understanding of complex physical processes in stellar interior
requires the introduction of free parameters that are calibrated to the
Sun. Therefore, using main sequence nearby stars with accurate fundamental
parameters to test the adequacy of extant stellar models is of paramount
importance to validate their range of applicability (Lebreton 2000), as we do
here.

The paper is organized as follows. In Section \ref{sample} we describe our
sample and in Section \ref{theory} we present our theoretical isochrones and
how they compare to observations. In Section \ref{method} we delve into the
derivation of the helium abundance for lower main sequence stars and in Section
\ref{notguilty} we carefully analyze how the results depend on the assumptions
made in stellar models.  The applicability range of our results is obtained by
comparing the prediction of the isochrones to a congruous set of main sequence
binaries (Section \ref{bitest}) and to empirical mass-luminosity relations
(Section \ref{masslum}). We suggest that an accurate mass-luminosity relation
for metal poor dwarfs could actually put constraints on their helium content.
In Section \ref{swp} we apply our method to derive masses and helium abundances
for a small set of planet host stars. We finally conclude in Section
\ref{conc}.

\section{Sample and Data selection}\label{sample}

Our sample stems from the 104 GK dwarfs for which we computed accurate
effective temperatures and bolometric luminosities via IRFM (Casagrande et
al. 2006).  For such stars accurate [$\alpha$/Fe] ratios and overall
metallicities [M/H] are available from spectroscopy as described in more
details in Casagrande et al (2006).  The main metallicity parameter in
theoretical models is in fact the total heavy-element abundance [M/H] and
neglecting the $\alpha$ enhancement can lead to erroneous or biased conclusions
(Gallart, Zoccali \& Aparicio 2005).  Once [M/H] is known, the metal mass
fraction $Z$ can be readily computed (see Appendix A).

We also paid special attention to removing unresolved double/multiple and
variable stars as described in full detail in Casagrande et al. (2006).

Some of the stars in the sample were too bright to have accurate 2MASS
photometry and therefore the IRFM could not be applied. However, such stars
have excellent $BV(RI)_C$ photometry so that from these colours it was possible
to recover the effective temperature and bolometric flux by means of the
multi-band calibrations given in Casagrande et al. (2006), homogeneously with
the rest of the sample. In this manner twenty-three single (or well separated
binary) and non variable stars were added to the sample. For these additional
stars $T_{eff}$ and $\mathcal{F}_{Bol}(\textrm{earth})$ have been estimated
averaging the values returned by the calibrations in all $BV(RI)_C$ bands. The
standard deviation resulting from the values obtained in different bands has
been adopted as a measure of the internal accuracy. The systematics due to the
adopted absolute calibration (Figure 12 in Casagrande et al. 2006) have then
been added to obtain the overall errors. When helium abundances for these
additional stars are computed (Section \ref{method}), the resulting values are
perfectly in line with those obtained for stars with fundamental parameters
obtained via IRFM.

We remark that an accurate estimate of the absolute luminosity (or magnitude)
of each star requires parallax accuracy at the level of a few percent. A
possible limitation could be the Lutz \& Kelker (1973) bias, however, as we
discuss in Appendix B, when limiting our sample to parallaxes better than 6\%
the bias is negligible compared to other uncertainties.  This requirement on
the parallaxes reduces our sample to 105 stars (see Figure \ref{best}).

We derive the helium content of our stars indirectly by comparing their
positions in the theoretical HR diagram with respect to a set of isochrones of
different helium and metallicity content (see Section \ref{method}). If
evolutionary effects have already taken place in stars, the comparison would be
age dependent (see Section \ref{theory}). However, for any reasonable
assumption about the stellar ages, one can safely assume that all stars fainter
than $M_V \sim 5.5$ are practically unaffected by evolution and lie close to
their Zero Age Main Sequence (ZAMS) location (e.g. Fernandes et al. 1996; Pagel
\& Portinari 1998; Jimenez et al. 2003). In terms of $M_{Bol}$ the threshold is
very similar (see figure 17 in Casagrande et al. 2006) so that we assume
$M_{Bol} \ge 5.4$ as a conservative estimate : this reduces the sample to 86 K
dwarfs.

For consistency with Casagrande et al. (2006), throughout the paper we assume
$M_{Bol,\odot}=4.74$ and $L_{\odot}=3.842 \times 10^{33}$~erg $s^{-1}$.

\section{Theory}\label{theory}

\subsection{Fine structure in the HR diagram: the broadening of the Lower Main 
Sequence}

Several effects are responsible for the observed width of the lower main
sequence : among the physical ones are chemical composition, evolution and
rotation whereas observational errors and undetected binarity among stars are
spurious ones.  We have carefully cleaned our sample from spurious effects
(Casagrande et al. 2006) so here we discuss only the physical ones related to
stellar structure.

The cut in absolute magnitudes adopted for our sample (Section \ref{sample})
ensures that our stars have masses below solar. The location of the main
sequence depends on the treatment of convection and the size of core convective
regions only for stars with $M > 1.1 M_{\odot}$ and on rotation for stars with
$M > 1.4 M_{\odot}$ (e.g. Fernandes et al. 1996) so these effects and the
related theoretical uncertainties are of no concern to us.

It is known that young and fast rotating K dwarfs might exhibit color anomalies
such as to alter their location on the HR diagram (e.g. Stauffer et
al. 2003). However, as we discuss in Section \ref{blue} this is not of concern
to us.

With typical masses $\sim 0.8\,M_{\odot}$ K dwarfs have lifetimes much longer
than the present age of the Galactic disk (e.g. Jimenez, Flynn \& Kotoneva
1998) and comparable to the present age of the Universe (Spergel et al. 2007)
so that evolutionary effects do not need to be taken into account.  For a given
metallicity $Z$, an increase of $Y$ makes a given mass on the isochrone hotter
and brighter so that the net result of varying $\Delta Y / \Delta Z$ is to
affect the broadening of the lower main sequence (see Figures \ref{best} and
\ref{tehe}). Such behaviour can be explained in terms of quasi-homology
relations (an increase of $Y$ in fact decreases the mean opacity and increases
the mean molecular weight e.g. Cox \& Giuli 1968; Fernandes et al. 1996) and it
has been exploited by Pagel \& Portinari (1998) and more recently by Jimenez et
al. (2003) to put constraints on the local $\Delta Y / \Delta Z$.

\subsection{Evolutionary tracks and isochrones}\label{Padova}

\begin{figure}
\begin{center}
\includegraphics[scale=0.42]{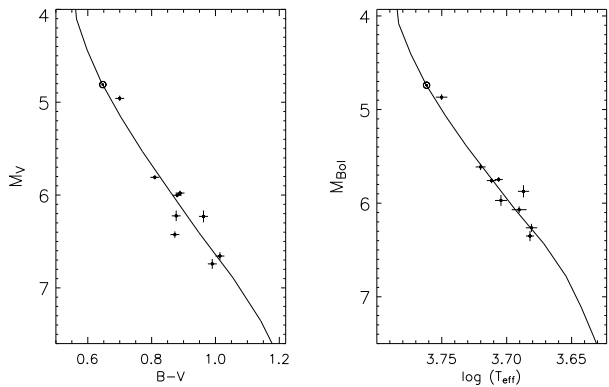}
\caption{{Comparison between 4.57~Gyr solar isochrone (age as determined from
    meteoritic measurements, Bahcall, Pinsonneault \& Wasserburg 1995) and our 
    sample stars in the [M/H] range $\pm 0.04$~dex around the solar value. 
    In order to have conservatively small errors we have used only stars with
    parallaxes better than 3\%. Transformations to plot the solar 
    isochrone in the observational plane as discussed in Section 
    \ref{obstheo}.
    Overplotted is also the Sun ($\odot$) for which we have adopted
    colours and temperatures from Casagrande et al. (2006). The difference 
    with the empirical $B-V$ of Holmberg et al. (2006) in unnoticeable.  
    Interestingly in the theoretical plane there is a much tighter agreement 
    between model and observations.}}
\label{solar}
\end{center}
\end{figure}

We have computed a series of stellar models, using the Padova code as in
Salasnich et al. (2000).  Since our sample includes only dwarfs, the
evolutionary tracks are limited to the main sequence phase only. We consider a
range of metallicities from $Z=0.0001$ to $Z=0.04$, and a range of $\Delta
Y/\Delta Z$ between 0 and 6. Metal abundance ratios are taken from Grevesse \&
Noels (1993). The solar metallicity is fixed to be $Z_\odot=0.017$, which is
the value preferred by Bertelli et al. (in preparation) to whom we refer for
all details. With this $Z_\odot$, the model that reproduces the present solar
radius and luminosity at an age of 4.6 Gyr was found to have $Y_\odot=0.263$,
and $\alpha_{\rm MLT}=1.68$. These numbers imply $(Z/X)_\odot=0.0236$ which is
slightly below the 0.0245 ratio quoted by Grevesse \& Noels (1993).  The
difference between the adopted $(Z/X)_\odot=0.0236$ and the Grevesse \& Noels
(1993) ratio simply implies a shift in the zero-point of the solar calibration
(from $Y_{\odot}=0.263$ to $0.289$) and it is of no concern as long as we are
interested in the study of a differential quantity such as $\Delta Y / \Delta
Z$. The adopted choice of $Y_{\odot}$ when combined with the latest $Y_P$ 
measurements (Section \ref{heres}) formally returns a $\Delta Y / \Delta Z$ in 
the range $0.7 - 0.9$. 
Our choice is to have a solar model that fits well the solar
position in the HR diagram (Figure \ref{solar}), 
and is by no means intended to be an accurate
re-calibration.  The latter would in fact be possible only by including atomic
diffusion in the solar model. Although in the Sun atomic diffusion is
efficient, spectroscopic observations of stars in Galactic globular clusters
and field halo stars (see discussion in Section \ref{DifSec}) point to a
drastically reduced efficiency of diffusion.  For this reason, the effect of
atomic diffusion is usually not included in the computation of large model
grids (e.g. Pietrinferni et al. 2004; VandenBerg, Bergbusch, Dowler 2006) and
the same approach is adopted here.

We also remark that the solar model is presently under profound revision after
the updates in the estimated solar abundances (Asplund, Grevesse \& Sauval,
2005) significantly diminish the agreement with helioseismology (e.g. Basu \&
Antia 2004; Antia \& Basu 2005; Delahaye \& Pinsonneault 2006).  The present
study however refers the ``classic'' solar model and metallicity as the
zero-point calibration, and since the method relies on differential effects
along the main sequence we do not expect that conclusions on $\Delta Y/\Delta
Z$, which is a {\it differential} quantity, are significantly affected. Also,
as already pointed out, the solar zero-point should be regarded as a
calibration parameter and not as an absolutely determined value so that strong
inferences on {\it absolute} values (among which $Y_P$) can hardly be
drawn. Therefore, as we will see later, small errors in this calibration
procedure, and especially in $Y_\odot$, are unlikely to affect our conclusions
on the $\Delta Y / \Delta Z$ obtained for our sample of stars. Nonetheless, we
caution on the hazardous comparison between the helium abundances deduced from
the isochrones and external constraints such as the primordial helium abundance
obtained with other techniques. In this case, in fact, differences in the
zero-point of $Y_{\odot}$ does not cancel out anymore. We will return on the
topic in Section \ref{heres}.

The tracks computed cover the mass range between 0.15 and 1.5 $M_\odot$, which
is far wider than needed for an analysis of our sample.  Regarding convection,
we adopt the same prescription as in Girardi et al. (2000): convective core
overshooting is assumed to occur for stars with $M>1.0 M_\odot$, with an
efficiency $\Lambda_c=M/M_\odot-1.0$ (see Bressan et al. 1993 for the
definition of $\Lambda_c$) that increases linearly with mass in the interval
from 1 to 1.5 $M_\odot$. Lower mass stars are computed assuming the classical
Schwarzschild criterion.

From these tracks, we can construct isochrones in the HR diagram for arbitrary
ages, and any intermediate $Y(Z)$ relation, via simple linear interpolations
within the grid of tracks.

\begin{figure*}
\begin{center}
\includegraphics[scale=0.90]{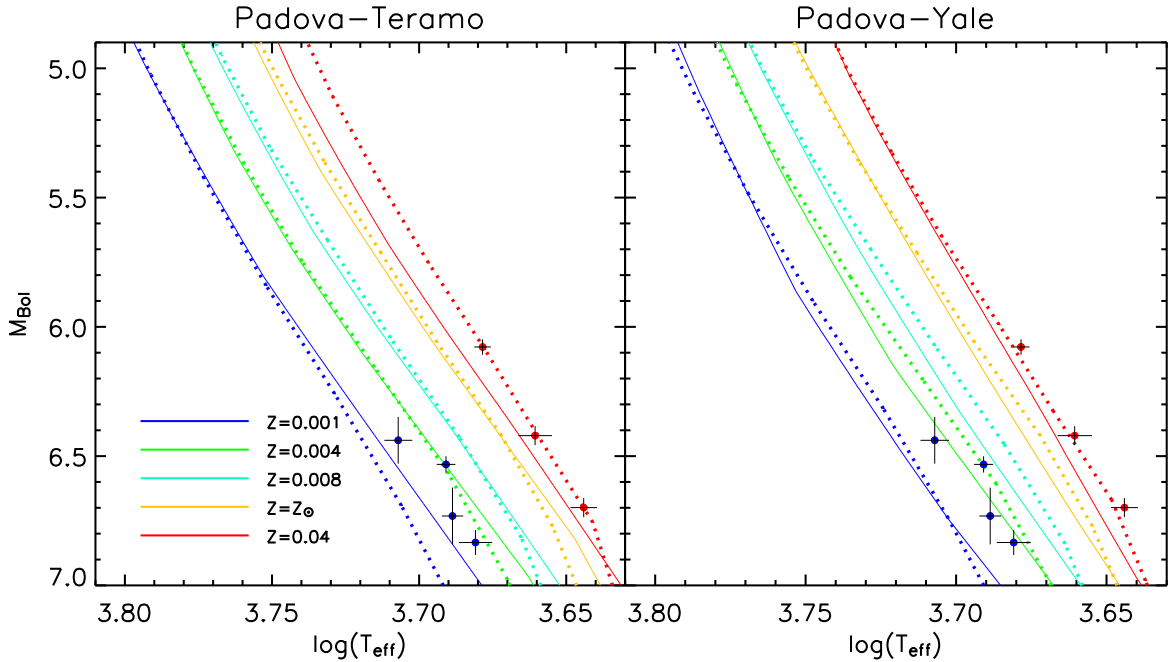}
\caption{Left panel: comparison between Padova (thick dotted) and Teramo (thin
  continuous) isochrones. Both sets are calibrated on the Sun and the
  comparison at other metallicities $Z$ is done for similar value of $\Delta Y
  / \Delta Z \sim 1.4$. Also shown for comparison are the most metal poor
  (blue) and metal rich (red) stars in our sample. Right panel: same as left
  panel, but with respect to Yale (thin continuous lines) isochrones ($\Delta Y
  / \Delta Z \sim 2.0 $)}
\label{pty}
\end{center}
\end{figure*}

Although our analysis is conducted using Padova isochrones, we have fully cross
checked (Section \ref{method}) the results with an updated set of the MacDonald
isochrones (Jimenez \& MacDonald 1996; Jimenez et al. 2003) computed for a
similar grid of values in $Y$ and $Z$.  We have also compared our isochrones
with other sets in the lower main sequence, namely the Teramo (Cordier et
al. 2007) and Yonsei--Yale (Demarque et al. 2004) isochrones.  The solar
isochrone of each set is calibrated to reproduce the current position of the
Sun in the HR diagram but the values of $Y_{\odot}$ and $Z_{\odot}$ are not
identical, because of the different prescriptions and input physics implemented
in various independent codes.  As we have already pointed out, what matters is
not the comparison between absolute values of $Y$ and $Z$. A meaningful
comparison can only be done between isochrones of similar $\Delta Y / \Delta Z$
with respect to a common calibration point i.e. with respect to the solar
isochrone. It is clear from Figure \ref{pty} that different set of isochrones
are in general in good agreement. The agreement between Padova and Yale
isochrones is outstanding throughout the entire range of metallicities and
$M_{Bol}$ covered in this study. The comparison with the Teramo isochrones is
also very good, except for the high metallicity isochrone that is significantly
hotter in the Teramo dataset. At lower metallicities, the agreement with Teramo
isochrones is always good except for the most metal poor and fainter stars in
our sample.  However, at luminosity higher than $M_{Bol}=6.0$ and below $\sim
4.5$ the agreement between Padova and Teramo isochrones is always outstanding.
A more detailed comparison is outside the purpose of the paper, but clearly our
results are not significantly affected by the particular set of isochrones used
(see also Figure \ref{yz}).

\subsection{Observational vs. theoretical plane}\label{obstheo}

Comparison between model isochrones and data is usually done in the
observational colour -- absolute magnitude HR diagram rather than in its
theoretical $T_{eff}$--$M_{Bol}$ counterpart. However, the observational plane
makes use of the information extracted only from a limited part of the entire
spectral energy distribution of a star (few thousands of \AA$\,$ in the case of
broad band colours, see also Section \ref{blue}).  Furthermore, for such a
comparison theoretical isochrones have to be converted into colours and
magnitudes via model atmospheres, introducing further model dependence and
uncertainties (e.g. Weiss \& Salaris 1999).

\begin{figure*}
\begin{center}
\includegraphics[scale=0.90]{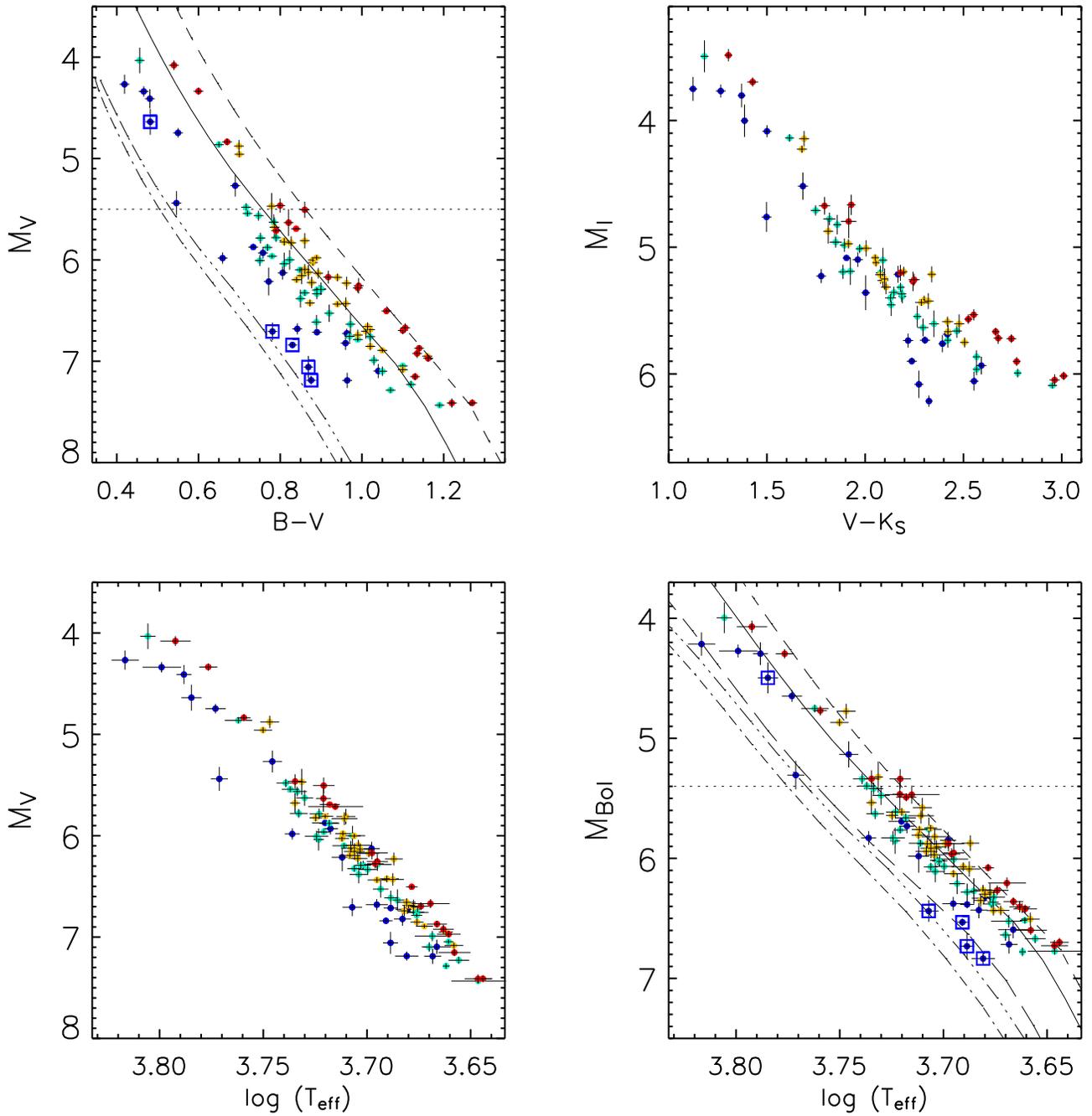}
\caption{{Comparison between observational, hybrid and theoretical planes.
    Parallax uncertainties are also included in error bars. Only stars with
    parallaxes better than 6\% are shown. For the second panel ($M_I,(V-K_S)$)
    only stars with accurate IR photometry (``j\_''$+$``h\_''$+$
    ``k\_msigcom''$< 0.10$) are shown. Points correspond to the sample stars in
    the range $Z < 0.007$ (blue), $0.007 \le Z < 0.014$ (cyan), $0.014 \le Z <
    0.022$ (yellow), $Z \ge 0.022$ (red). Squares are intended to highlight
    stars with $Z \sim 0.001$ to facilitate comparison with metal
    poor isochrones.  Overplotted are 1~Gyr isochrones of metallicity $Z=0.001$
    (dot dashed), 0.017 (continuous) and 0.040 (dashed) under the standard
    assumption of $\Delta Y/\Delta Z = 2$. The lower and upper $Z$ value
    roughly bracket the metallicity of our sample. Also shown for comparison
    (triple dot dashed) is an extremely helium poor isochrone ($Z=0.001$ and
    $Y=0.167$). In the fourth panel is also plotted an isochrone with $Z=0.004$
    and $Y=0.250$ (long dashed line).}}
\label{best}
\end{center}
\end{figure*}

Sometimes the hybrid $T_{eff}$ -- absolute magnitude plane is used although it
has almost the same limitations as the observational one (the computation of
absolute magnitudes from stellar models still requires the use of spectral
libraries).  In addition $T_{eff}$ is rarely empirically and consistently
determined for all the stars, more often depending on the adopted
colour--temperature transformation or resulting from a collection of
inhomogeneous sources in literature.  As a result, a single theoretical
isochrone produces different loci in the color--magnitude diagram when
different color-temperature relations are applied (e.g. Pinsonneault et
al. 2004).

Working in the purely theoretical plane has many advantages.  First, it is
possible to directly compare observations to physical quantities such as
temperature and luminosity (or equivalently $M_{Bol}$) predicted from stellar
models.  Secondly, for the purpose of this work, the effects of helium are
highlighted in the theoretical plane, whereas in the widely used $(B-V)$ vs
$M_V$ plane they are partly degenerate with the dependence of the colour index
on metallicity (Castellani, Degl'Innocenti \& Marconi 1999).  Up to now, the
drawback of working in the theoretical plane was that for a given set of stars,
very rarely in literature homogeneous and accurate bolometric corrections and
effective temperatures were available for large samples. However, for all our
stars we have homogeneously derived $T_{eff}$ and bolometric luminosity from
accurate multi--band photometry and basing on the IRFM (Section \ref{sample}).
In the case of K dwarfs $\sim$80\% of the total luminosity is directly observed
so that the dependence on model atmosphere is minimal\footnote{We further point
out that even if model atmospheres are computed with a standard helium content,
the spectral energy distribution is largely insensitive to the helium abundance
(e.g. Peterson \& Carney 1979; Pinsonneault et al. 2004).  A $\Delta Y \sim
0.10$ in model atmospheres changes synthetic magnitudes by $\Delta m \sim 0.01$
(Girardi et al. 2007). Our implementation of the IRFM relies on multi-band
photometry and uses model atmospheres to estimate the missing flux only to a
limited extent (few tens of percent). Our results are thus unaffected by this
uncertainty.}.

The first feature that appears from the comparison between the observational
$(B-V)$ and the theoretical HR diagram (Figure \ref{best}) is how in the
$T_{eff} - M_{Bol}$ plane the separation between metal poor and metal rich
stars is not as neat as in the observational $(B-V)$ counterpart.  This was
already noticed by Perrin et al. (1977) and Lebreton et al. (1999) and reflects
the sensitivity of the $B,V$ colour indices to metallicity. To ensure this is
not an artifact due to our temperature and/or luminosity scale, we also show a
plot in the observational $M_I$ versus $(V-K_S)$ plane. $(V-K_S)$ is in fact
and excellent temperature indicator with negligible dependence on metallicity
and $M_I$ faithfully traces $M_{Bol}$ (Casagrande et al. 2006). The reduced
separation between metal poor and metal rich stars is confirmed.

To quantify how strong is the effect of metallicity in the $(B-V)$ plane, we 
have drawn theoretical isochrones in this plane, too. The transformations 
to convert theoretical isochrones into the observational plane have been 
obtained by fitting the following formulae to the stars in Casagrande et al. 
(2006) (their figure 17):
\begin{displaymath}
BC = a_0 + a_1 T_{eff} + a_2 T_{eff}^2 + a_3 T_{eff} \textrm{[M/H]} + a_4
\textrm{[M/H]} 
\end{displaymath} 
\begin{equation}\label{eqBC}
\phantom{BC =} + a_5 \textrm{[M/H]}^2
\end{equation} 
where $BC$ is the bolometric correction (from which $M_V = M_{Bol}-BC$) and: 
\begin{displaymath}
B-V = b_0 + b_1 T_{eff} + b_2 T_{eff}^2 + b_3 T_{eff} \textrm{[M/H]} + b_4
\textrm{[M/H]} 
\end{displaymath} 
\begin{equation}\label{eqBV}
\phantom{B-V =} + b_5 \textrm{[M/H]}^2.
\end{equation} 
Both transformations are accurate to 0.02~mag and the coefficients are given
in Table \ref{coef}.
\begin{table*}
\centering
\caption{Coefficients $i=a,b$ for equation (\ref{eqBC}) and (\ref{eqBV}) in
  the temperature range $4300 < T_{eff} < 6700$~K.}
\label{coef}
\begin{tabular}{lcccccc}
\hline
       & $i_0$    & $i_1$   & $i_2$  & $i_3$ & $i_4$    & $i_5$ \\
\hline
$BC$  & $-7.67005$ & 0.00248 & $-2$E$-7$ & 9E$-5$ & $-0.44985$ & 0.00672 \\
$B-V$ &  4.96959 & $-0.00120$ & $8$E$-8$ & $-3$E$-5$  &  0.29112 & 0.03581 \\
\hline
\end{tabular}
%
\end{table*}

With this approach the comparison between isochrones and sample stars in the
two planes does not depend on model atmospheres, since we use empirical
conversions derived from the same sample stars. Notice though that the
empirical conversions of Casagrande et al. (2006) show good agreement with
theoretical ones from e.g. Kurucz or MARCS model atmospheres.

From Figure \ref{best} is clear that for metallicity around and above the
solar one, isochrones with $\Delta Y / \Delta Z=2$ are in overall good 
agreement with the data. On the contrary, a clear discrepancy appears
for metal poor stars where the standard assumption $\Delta Y / \Delta Z=2$
returns isochrones that are too hot. 

To achieve a match between the metal poor stars and the isochrone, we need to
decrease the corresponding helium abundance down to $Y=0.167$, {\it well below}
the primordial value expected from Big Bang nucleosynthesis. Even with such a
radically low helium abundance the discrepancy is persistent. An alternative to
reduced helium in the stars is to use a theoretical isochrone with a more
orthodox helium abundance ($Y=0.250$, (fourth panel in Figure \ref{best}) but
with a metallicity ($Z=0.004$) higher by $\sim 0.6$~dex (with respect to
$Z=0.001$), a very large change in metallicity content indeed (and discussed in
detail in Section \ref{nlte}).  This comparison qualitatively illustrates how
discrepant the lower metallicity stars are compared to models. This result is
discussed in the detailed analysis of Section \ref{method}.

\section{Helium abundance and mass from theoretical isochrones}\label{method}

In the previous Section we have shown that the most suitable place to estimate
the effects of the helium content on the lower main sequence is the theoretical
$T_{eff} - M_{Bol}$ plane. In this section we follow this by fitting to each
star the most appropriate isochrone; the metallicity $Z$ of each star is known
from its spectroscopic measurement (see also Appendix A) and we thereby
determine its helium content $Y$. We thus differ from previous works which
--by means of different techniques-- focused on the overall comparison
between theoretical predictions and observations along the lower main sequence
(Fernandes et al. 1996; Pagel \& Portinari 1998; Lebreton et al. 1999; Jimenez
et al. 2003). Our approach also avoids any assumption about the existence of a
constant (linear) helium-to-metal enrichment rate $\Delta Y / \Delta Z$.  As we
discuss later, such a constant ratio may well apply for metallicity around and
above the solar one but at lower metallicity the situation is far less clear.

To first order, the position of a star in the HR diagram (i.e. its $M_{bol}$
and $T_{eff}$) depends on its chemical composition (i.e. $Y$ and $Z$), mass and
age (e.g. Fernandes \& Santos 2004). The broadening of the lower main sequence
however is independent of the age, meaning that at increasing age low mass
stars move --very slowly-- on the HR diagram along a direction that is
roughly parallel to the main sequence.  Therefore, even though a correct choice
of the age is important to properly estimate the mass of lower main sequence
stars, their helium content does not depend on it. Since in the present
investigation we are primarily interested in determining the helium abundance
of our sample stars, an accurate estimate of the age of our stars is not
required. As we will see, changing the age of the isochrones by a large amount
barely changes the derived helium content of low mass stars.

Besides mass, age, $Y$ and $Z$, there are other physical parameters used to
describe the stellar interiors in the models. Of particular significance is the
mixing length parameter $\alpha_{\rm MLT}$.  Effective temperature, luminosity,
mass, age and metallicity are known with great accuracy for the Sun, so that a
stellar model can be made to fit the Sun by adjusting only two free parameters
($Y$ and $\alpha_{\rm MLT}$).  For stars other than the Sun, this procedure has
been done by calibrating stellar models to a few nearby visual binary stars
(Fernandes et al. 1998; Lebreton et al. 2001; Fernandes, Morel \& Lebreton
2002; Pinsonneault et al.  2003) with particular attention to the $\alpha$~Cen
system (see discussion in Section \ref{notguilty} and \ref{bitest}).
Unfortunately, for these binary systems the uncertainties in the fundamental
physical parameters required to calibrate stellar models are rather large as
compared to the Sun so that in the final set of calibration parameters there is
a certain degeneracy.  In this respect our approach is much more
straightforward since the adjustable parameters ($Y$ and $\alpha_{\rm MLT}$)
are strictly calibrated on the Sun (Figure \ref{solar}).  Such calibrated model
is then used to compute a large grid of tracks with different metallicities
($Z$) and helium ($Y$) content from which isochrones are constructed. Our grid
is used to deduce helium abundances and masses for field stars and the results
are then validated checking our procedure with a congruous number of binaries
(Section \ref{bitest}).

\subsection{Method}\label{metodo}

\begin{figure*}
\begin{center}
\includegraphics[scale=0.46]{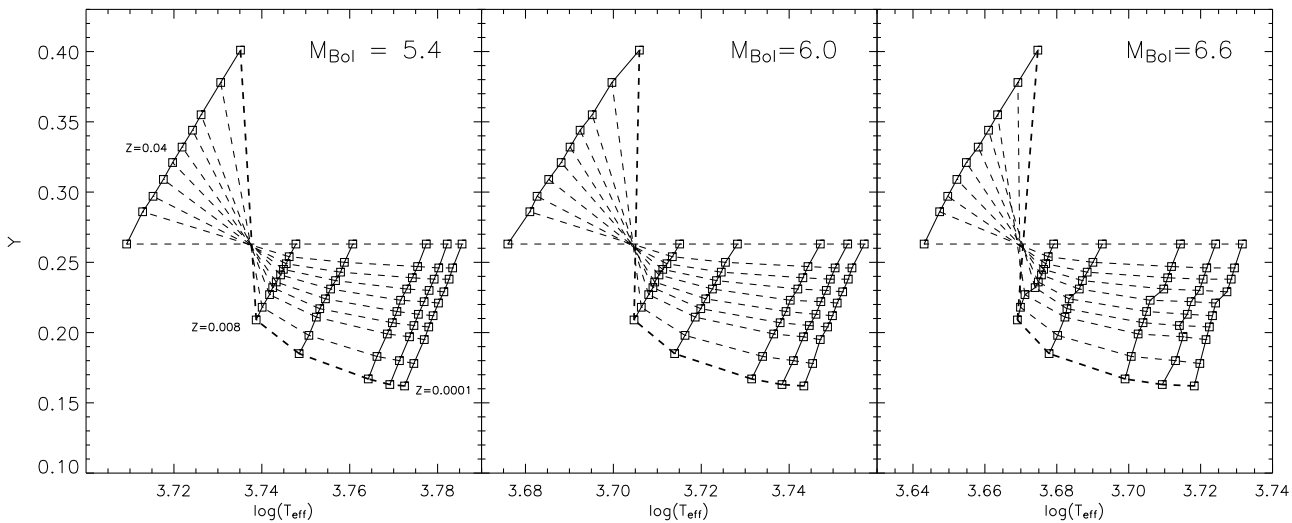}
\caption{$\log (T_{eff})$--$Y$ relation for different isochrones at three given
  values of $M_{Bol}$. Continuous lines connect squares of equal metallicity 
  $Z$. The thick dash line refers to $\Delta Y / \Delta Z = 6$, the thin dash 
  line to $\Delta Y / \Delta Z = 0$. Others dash lines refer to intermediate
  values. Notice that different $\Delta Y / \Delta Z$ pivot around the
  temperature of our reference solar isochrone for the given $M_{Bol}$.}
\label{tehe}
\end{center}
\end{figure*}

As we have discussed in Section \ref{theory} at any given $M_{Bol}$, for a
given metallicity $Z$, an increase of $Y$ translates into an increase of
$T_{eff}$ of the isochrone.  This can be easily explained in term of
quasi-homology relations and our grid of isochrones clearly confirms this
behaviour (Figure \ref{tehe}).

\begin{figure*}
\begin{center}
\includegraphics[scale=0.46]{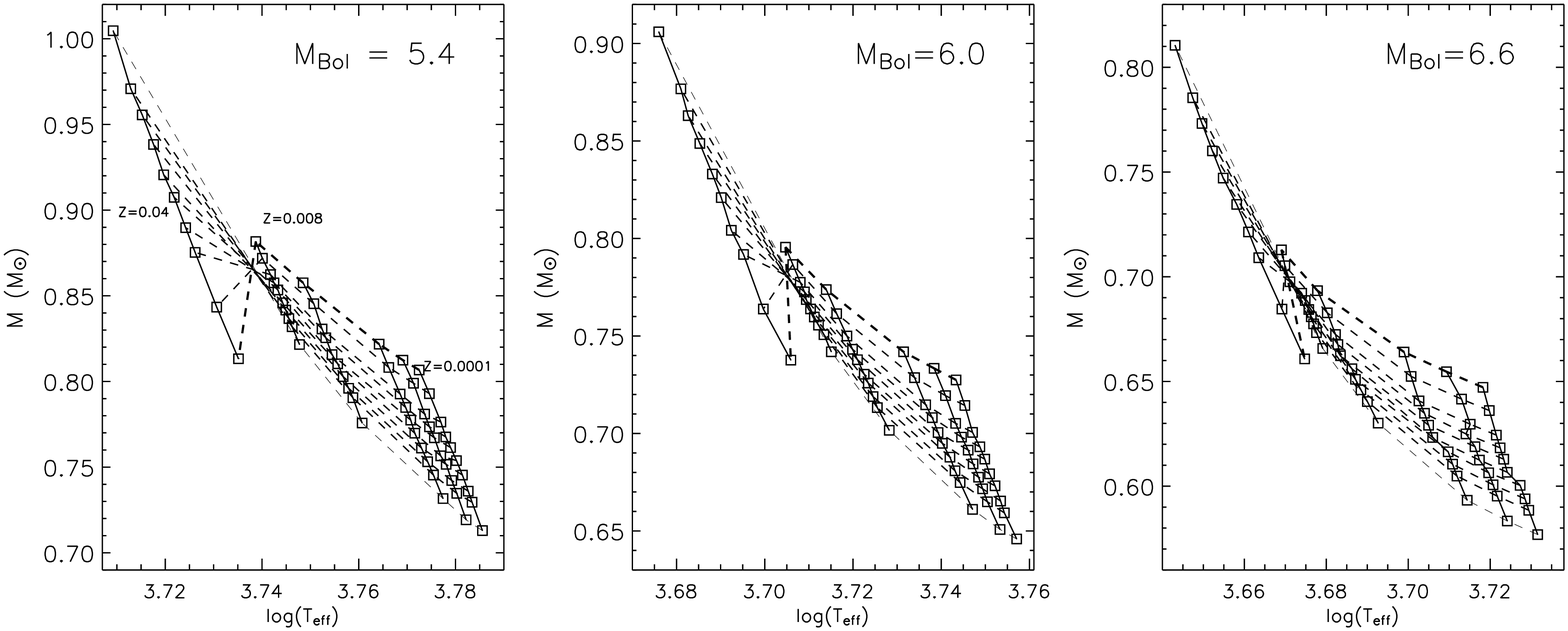}
\caption{$\log (T_{eff})$--$\textrm{Mass}$ relation 
  for different isochrones at three given values of $M_{Bol}$. Same
  prescriptions as in the previous figure.}
\label{tema}
\end{center}
\end{figure*}

\begin{figure*}
\begin{center}
\includegraphics[scale=0.46]{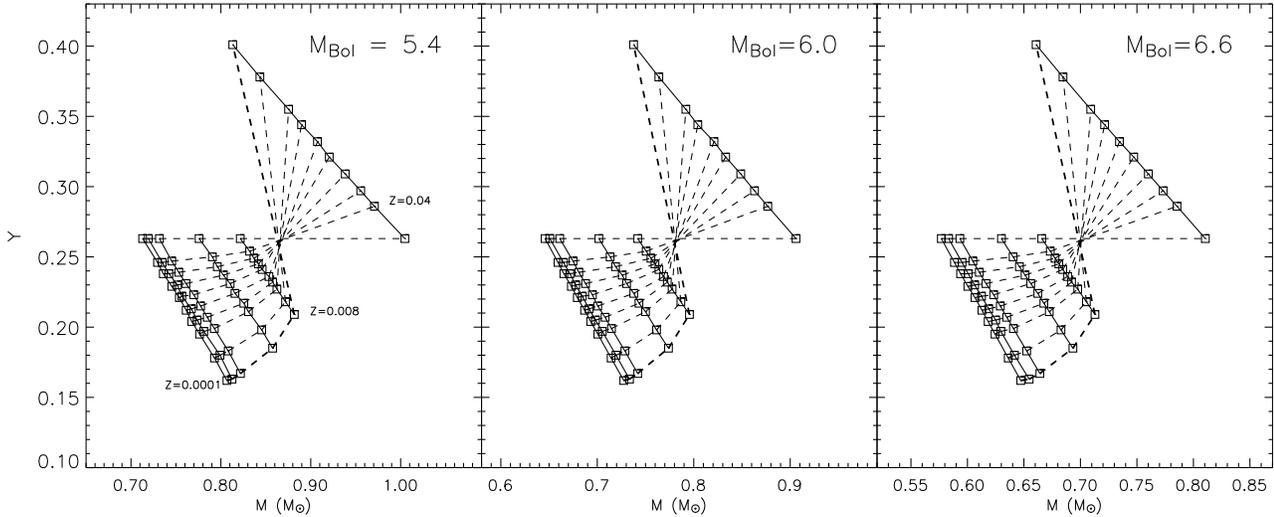}
\caption{$\textrm{Mass}$--$Y$ relation for different
  isochrones at three given values of $M_{Bol}$. Same prescriptions as in 
  Figure \ref{tehe}.}
\label{hema}
\end{center}
\end{figure*}

Since for our 86 K dwarfs $M_{Bol}, T_{eff}$ and $Z$ are known (Section
\ref{sample}) it is possible to infer the helium fraction $Y$ with a simple
interpolation over grids of the kind in Figure \ref{tehe}.  Analogous grids
exist between $T_{eff}$ and mass (Figure \ref{tema}) and $Y$ and mass (Figure
\ref{hema}) so that from the isochrones it is also possible to infer the mass
once the age is chosen.

We use 5~Gyr old isochrones, half of the age of the disk (e.g. Jimenez et
al. 1998) and consistent with the age of nearby solar-type stars (Henry et
al. 1996). Since the stars are virtually unevolved, the choice of age is not
critical. We have tested the difference if 1~Gyr and 10~Gyr old isochrones are
used instead. With respect to the adopted choice of 5~Gyr, younger (older)
isochrones yield helium abundances lower (higher) by $\sim 0.005$ and masses
larger (smaller) by $\sim 0.03 M_{\odot}$, the biggest differences occurring at
the higher masses covered in this study.  As we show in Section \ref{heres},
such differences in helium abundance are considerably smaller than those
stemming from the uncertainties in parallax, $T_{eff}, M_{Bol}, Z$.  Recent
studies of dwarfs stars in the Solar Neighbourhood do suggest a typical age of
about 5~Gyr, with considerable scatter (Reid et al. 2007). While using 5~Gyr
old isochrones might not be the most accurate choice for any given star, the
trend defined by our masses (Section \ref{masslum}) should be on average
correct. At the very least, with such a choice on the age, the masses of the
few available nearby binaries are almost all recovered with good accuracy
(Section \ref{bitest}).

Some of the stars turn out to be outside the grid, meaning their inferred
helium content lies outside the range covered by the isochrones. Since the grid
is very regular and linear relations are also expected from quasi-homology, we
have used a linear fit to extrapolate the helium content and the mass.  As a
consistency check we have also adopted another approach by fitting a second
order polynomial between the helium content $Y$ and $T_{eff}, M_{Bol}$ and
$Z$. The helium abundances obtained with the two methods are identical to
better than 0.01 meaning that the adopted extrapolation procedure has a
negligible effect on the overall results. As a further test we have also used
the MacDonald isochrones and the deduced helium abundances are always in very
good agreement with those obtained with Padova ones, again confirming that our
results do not depend significantly on the particular set of isochrones used.

\subsection{Results}\label{heres}

\begin{figure*}
\begin{center}
\includegraphics[scale=0.70]{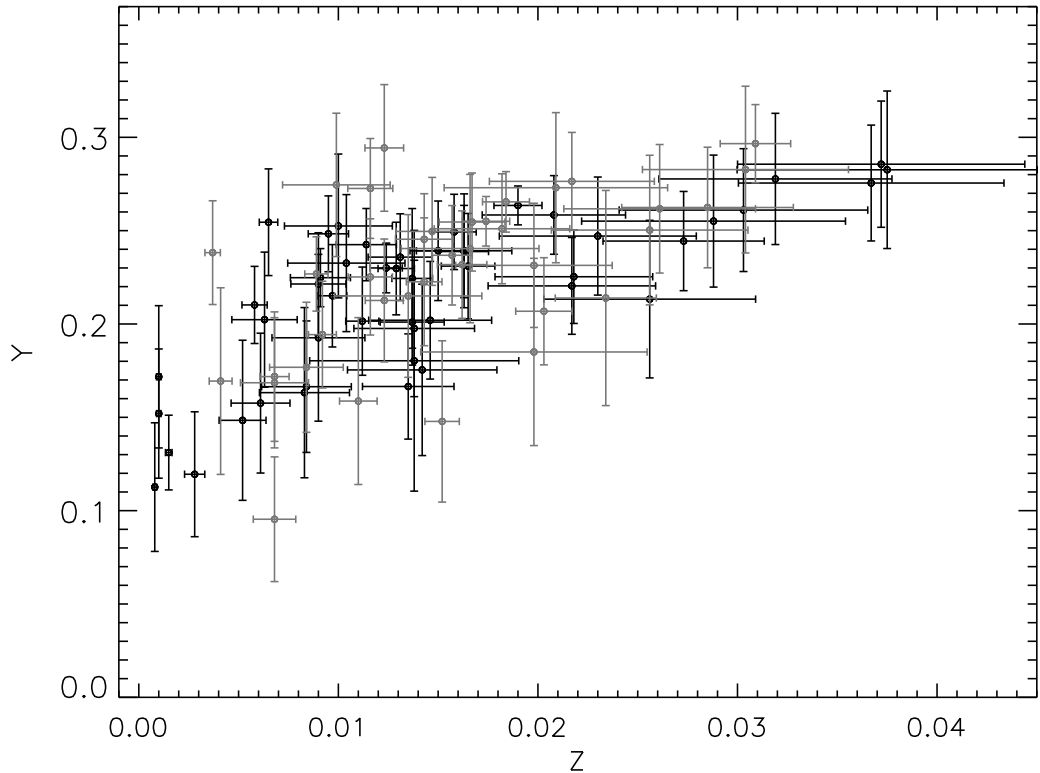}
\caption{{Helium ($Y$) to metal ($Z$) enrichment factor for our sample of
    stars. Grey points are for stars with $M_{Bol} \le 6.0$, where
    different sets of isochrones are formally identical at low metallicities
    (see Figure \ref{pty}). The
    occurrence of low helium abundances is confirmed. 
    Error bars from MonteCarlo simulation according to the
    prescription given in the text.}}
\label{yz}
\end{center}
\end{figure*}

The behaviour of $Y$ with $Z$ is shown in Figure \ref{yz}. Error bars for the
helium abundances have been obtained via MonteCarlo simulations, assigning each
time values in parallaxes, metallicities, temperatures and bolometric
luminosities with a normal distribution centered on the observed values and a
standard deviation equal to the errors of the aforementioned quantities. Since
the age chosen for the isochrones has negligible effects on the helium
abundances (Section \ref{metodo}), we have not accounted for any dependence on
the age, that we keep fixed at 5~Gyr.  The MonteCarlo returns typical errors of
order 0.03 in $Y$ and $0.03 - 0.04 \, M_{\odot}$ in mass.  In the worst case
scenario a large variation of the age can introduce an error in mass of similar
size (Section \ref{metodo}), increasing therefore the final uncertainty to
$0.04-0.06 \, M_{\odot}$.

It is clear from Figure \ref{yz} that the helium-to-metal enrichment ratio is
roughly linear for metallicities around and above the solar one.  A linear fit
in this range is in fact able to recover within the errors the solar
calibration value, although the formal extrapolated primordial $Y_{P}$ is
underestimated with respect to Big Bang Nucleosynthetic estimates (Table
\ref{ratio}). A plot of this kind was also done by Ribas et al. (2000) who
fitted a large grid of stellar evolutionary models to a sample of detached
double-lined eclipsing binaries with accurately measured absolute dimensions
and effective temperatures. With this approach they were able to simultaneously
determine $Z$ and $Y$ (both kept as free parameters) for 28 systems. Despite
the very different approach and the fact they preferentially studied evolved
stars, the comparison with their work is very telling. Their models were
calibrated with slightly different $Z_{\odot}$ and $Y_{\odot}$ (Claret 1995)
thus implying a shift in the zero-point of the $\Delta Y / \Delta Z$ plot, but
their slope is very similar to ours, also considering their sample was limited
to metallicities somewhat higher than we have here.  At their lowest $Z$ the
scatter seems to increase and low helium abundances appear, but unfortunately
there are too few stars in common to draw any firm conclusion.

In our Figure \ref{yz} a puzzling turnover in the helium content appears going
to lower metallicities, with a break around $Z=0.013$, reflecting what was
qualitatively expected given the premises discussed in Section
\ref{obstheo}. Also, at lower metallicities the scatter in the data is larger.
Such low helium abundances are clearly at odds with the latest primordial
helium measurements from H~II regions that range from $Y_P=0.2472$ to
$Y_P=0.2516$ (Peimbert, Luridiana \& Peimbert 2007; Izotov, Thuan \&
Stasinska 2007). Within present day accuracy, CMB data alone constrain the
primordial helium mass fraction only weakly $0.160 < Y_{P} < 0.501$ (Trotta \&
Hansen 2004). What it is actually measured in CMB data is the baryon-to-photon
ratio; once Standard Big Bang Nucleosynthesis is assumed a formally precise
$Y_P=0.24815$ can be calculated (Spergel et al. 2007).

Although the discrepancy with the primordial helium abundance is significant,
we stress that the solar $Y_{\odot}$ is a calibration parameter in stellar
tracks (Section \ref{Padova}).  Therefore, a meaningful comparison can only be
done for abundances obtained with the same technique. For this reason we
expect our conclusions on $\Delta Y / \Delta Z$ to be rather robust. It is
clear that Figure \ref{yz} casts some doubts whether the hypothesis of a linear
trend with constant $\Delta Y / \Delta Z$ ratio is necessarily true over the
entire $Z$ range.  Helium--to--metal enrichment factor determinations in
literature have often made such an assumption for the sake of simplicity: in
our case the ratio changes from 4 when all stars are considered to
approximately 2 when only metallicity around and above the solar one are used :
this may partly explain the very different measured values of $\Delta Y /
\Delta Z$ often reported in literature.

In the case of our isochrone fitting procedure, small changes in the zero-point
of $Y_{\odot}$ (see Section \ref{Padova}) could partly attenuate the
discrepancy with the primordial helium abundance measured with other
techniques.  Even so, the significantly low helium abundances found for metal
poor stars are very difficult to explain.  The need for sub-primordial helium
abundances to fit the lowest metallicity stars in the Solar Neighbourhood with
extant stellar models was already highlighted by Lebreton et al. (1999).  Such
a large discrepancy is evidently a challenge for stellar modeling and/or basic
stellar data.  In the next section we carefully discuss to which extent such a
large discrepancy can be reduced.

\begin{table}
\centering
\caption{Linear fit with errors on both axis for the data in Figure
  \ref{yz}. $Y_{\odot}$ is the solar value recovered from the fit, to be
  compared with the value of 0.263 used to calibrate our solar model.}
\label{ratio}
\begin{tabular}{cccc}
\hline
             & $\frac{\Delta Y}{\Delta Z}$ &  $Y_P$         & $Y_{\odot}$    \\
\hline
$Z \ge 0.013$ &  $3.2 \pm 0.9$            & $0.18 \pm 0.02$ & $0.24 \pm 0.02$\\
$Z \ge 0.015$ &  $2.1 \pm 0.9$            & $0.21 \pm 0.02$ & $0.24 \pm 0.02$\\
$Z \ge 0.018$ &  $1.8 \pm 1.1$            & $0.22 \pm 0.02$ & $0.25 \pm 0.03$\\
\hline
\end{tabular}
\end{table}

\section{Searching for partners in crime}\label{notguilty}

We discuss next various theoretical and observational uncertainties that could
affect our results. 

First, we have searched for any correlation between helium content and various
parameters other than metallicity to highlight possible spurious trends in the
models; then we focus on some open problems related to stellar models and
finally to the adequacy of the adopted metallicity, temperature and luminosity
scales.

\subsection{High rotational velocities}\label{blue}

Rotation is usually neglected in standard stellar evolutionary codes and the
license of this choice holds as long as the stars studied are not significantly
affected by rotation themselves.
 
Young stars have usually high rotational velocities that might deposit large
amount of non-radiative heating into their outer layers, thus significantly
affecting the observed colours.  The effect is well documented in Pleiades'
dwarfs (e.g. Jones 1972; Stauffer et al. 2003) and there is evidence it might
also occurs in other young clusters (An et al. 2007).  In the Pleiades this
phenomenon is termed ``blue K dwarfs'', with the stars lying nearly half a
magnitude below the main-sequence isochrone in the $(B-V)$ vs $M_V$
plane. However, in $(V-K)$ vs $M_V$ the situation is reversed, with Pleiades K
dwarfs now systematically brighter (Stauffer et al. 2003).  
Opposite behaviour in different colour indices cautions us to the risk of
introducing biases when a certain photometric system --that might highlight or
shadow certain features-- is used. Our use here of physical quantities such as
bolometric luminosities and effective temperatures is clearly safer.

We do not expect high rotational velocities in our sample of stars since
rapidly rotating K dwarfs also exhibit photometric variability with period of
the order of few hours and $V$ magnitude amplitudes up to $\sim 0.15$
magnitudes (van Leeuwen et al. 1986, 1987), more slowly rotating K dwarfs being
generally less photometrically variable (e.g. Stauffer \& Hartmann 1987;
Terndrup et al. 2000). As described in more detail in Casagrande et al. (2006)
our sample has been cleaned of variable stars to a high accuracy level so that
we do not expect any rapidly rotating star in our sample.  For the sake of
completeness, we have searched for $V_{rot} \sin (i)$ measurements from the
comprehensive catalog of Glebocki \& Stawikowski (2000). We found measurements
for 45 out of 86 K dwarfs in our sample, taking the mean value when multiple
measurements were available.  As expected, all stars have low rotational
velocities. Also, it is clear from Figure \ref{dep} that the derived helium
abundances are independent of $V_{rot} \sin (i)$.

\subsection{Evolutionary effects}

\begin{figure*}
\begin{center}
\includegraphics[scale=0.70]{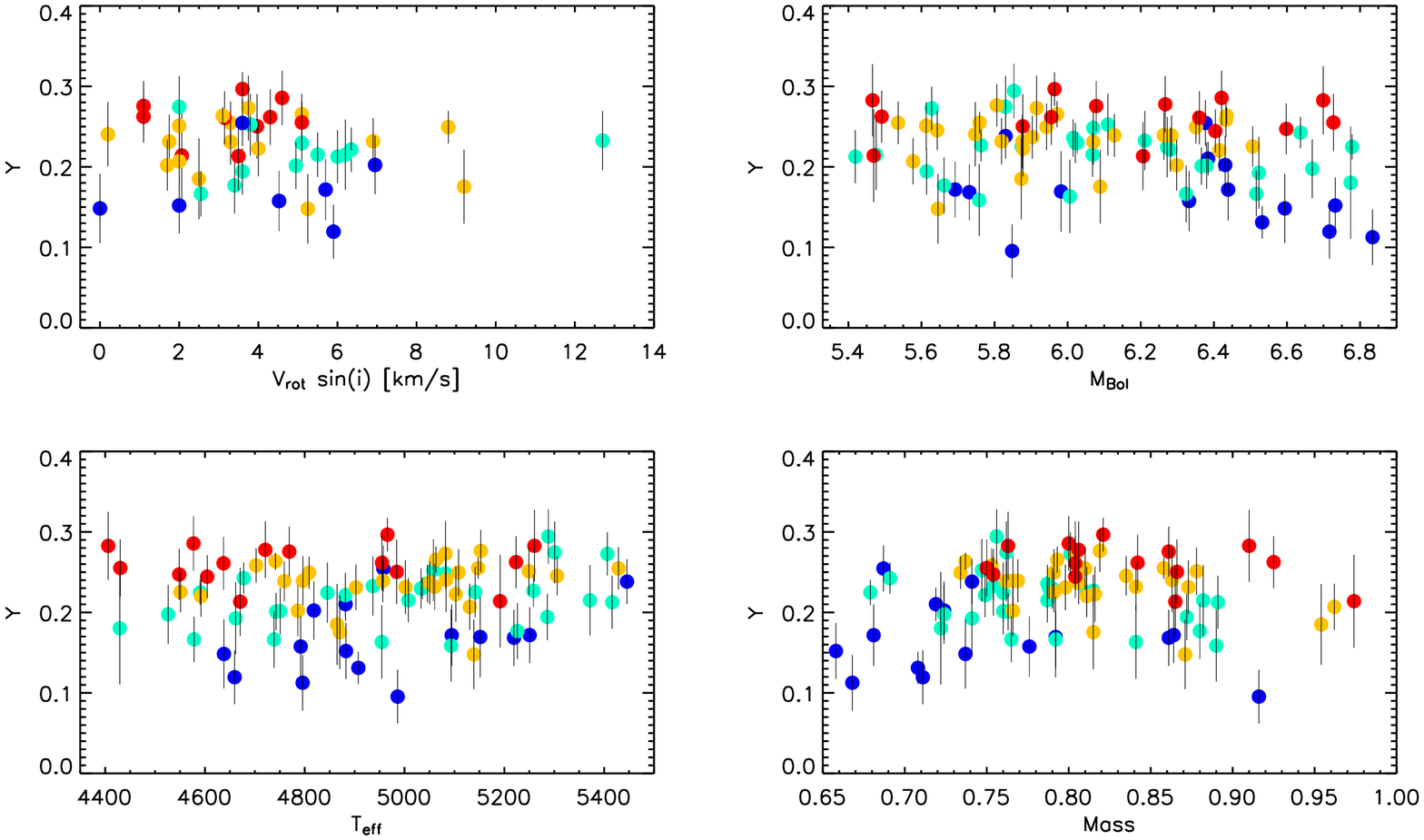}
\caption{{Helium content deduced from the isochrones 
    as function of $V_{rot} \sin(i)$ (available for 45 K dwarfs), $M_{Bol}$, 
    $T_{eff}$ and Mass for all our 86 K dwarfs. Stars are divided in the same 
    metallicity bins and colours as in Figure \ref{best}. 
    No obvious dependence appears and low $Y$ values are practically present
    throughout the entire range of parameters covered by our stars.}}
\label{dep}
\end{center}
\end{figure*}

Evolutionary effects are particular important, since low helium abundances
could results from the attempt of fitting isochrones to stars that have already
departed from their ZAMS and are thus brighter: in this case we expect a
correlation between $Y$ and $M_{Bol}$. Figure \ref{dep} shows the helium
content as function of $M_{Bol}$, $T_{eff}$ and mass (deduced from the
isochrones). Since these plots have already built-in the $Y-Z$ correlation that
could mask or counterbalance other correlations, we have divided the sample
into four metallicity bins to disentangle the underlying $Y(Z)$ correlation
from the others. No significant other correlation appears, besides the expected
split between metallicity bins.  At low metallicities, helium abundances below
$Y=0.20$ are practically present for any value of $M_{Bol}$, $T_{eff}$ and
mass.

The absence of any obvious trend is a posteriori confirmation of the adequacy
of the adopted evolutionary cut $M_{Bol} \ge 5.4$. Figure \ref{dep} shows that
most of our stars have masses (deduced from the isochrones) below $0.85
M_{\odot}$ at low $Z$.  Studies of globular clusters in the Milky Way also
confirm that metal-poor stars below $0.80 - 0.85 M_{\odot}$ have not yet
reached the turn-off, the exact value depending on the age and the underlying
details of the isochrones used to fit a globular cluster (e.g. Chaboyer et
al. 2001; Morel \& Baglin 1999). For higher metallicity the turn-off mass is
also higher, making evolutionary effects even less likely in such stars in our
sample.

\subsection{Shortcomings in stellar models}

Despite the steady improvement in modeling stellar structure and evolution,
there are still shortcomings in the theory that require the introduction of
adjustable parameters, typically calibrated on the Sun.

Our model is calibrated on the Sun, for an assumed $Z_{\odot}=0.017$, by
adjusting the helium content and the mixing-length in order to match its
present age, radius and luminosity (Section \ref{Padova}).  As we already
pointed out, the $Y_{\odot}$ value must be intended as the zero-point of our
calibrated model and not as the absolute value of the solar helium content.
Helioseismology does in fact return a lower helium content but including
diffusion in the model helps to reduce such a difference (see Section
\ref{DifSec}).  The difference between the present helium value derived from
seismology and the initial value obtained from the calibration provides a
constraint on the input physics of the model.

The fact that we are working with stars that are only slightly cooler and
fainter than the Sun should ensure that we are studying a region of the HR
diagram where models, at least for metallicities around the solar one, are well
calibrated. Our solar isochrone is in fact in outstanding agreement with a
sample of solar metallicity stars (Figure \ref{solar}).

\subsubsection{Mixing-length}\label{mixi}

The universality of the mixing-length value is an open question.  The analysis
of binaries in the Hyades has recently lead Lebreton et al.  (2001) and Yildiz
et al. (2006) to conclude that the mixing-length increases with stellar mass.
Similar conclusions were also drawn by Morel et al. (2000a) and Lastennet et
al. (2003) based on the study of the binary systems $\iota$~Peg and UV Piscium,
respectively.  These results are opposite to the theoretical expectation from
hydrodynamical simulations of convection (Ludwig, Freytag \& Steffen 1999;
Trampedach et al.  1999).  Detailed calibration of stellar models on the
$\alpha$~Cen system have returned discordant conclusions about the universality
of the mixing length parameter (e.g. Noels et al. 1991; Edmonds et al. 1992;
Neuforge 1993; Fernandes \& Neuforge 1995; Morel et al. 2000b; Guenther \&
Demarque 2000).  The latest model calibrations on the $\alpha$~Cen system
making use of seismic constraints favor a mixing-length that increases going to
lower mass (Eggenberger et al. 2004; Miglio \& Montalb\'an 2005) and therefore
in agreement with the theoretical expectations.  The discordant conclusions
drawn from all these studies probably reflect the many observational
uncertainties (order of magnitudes larger than for the Sun) in the input
parameters of the models.  These results suggest that at this stage a clear
relation between mass and mixing-length is premature, either because
uncertainties in the input parameters can overshadow shortcomings in the
mixing-length theory itself, or because a dispersion of mixing-length at a
given mass or even a time dependence of the mixing-length (Yildiz 2007) could
well be possible. Therefore, assuming the solar mixing length is currently the
safest choice. Systematic trends in mixing length are anyways overwhelmed by
observational uncertainties.

As regards the dependence on the metallicity, the fact that all globular
clusters can be fit with the same value for the mixing length parameter
supports the assumption that it does not depend on $Z$, although such a
conclusion is obtained studying giant branch stars only (e.g. Jimenez et
al. 1996; Palmieri et al. 2002; Ferraro et al. 2006).  Concerning the
particular region of the HR diagram we are going to investigate, models
computed with the solar mixing-length reproduce the slope of the main sequence
of young open clusters quite well (VandenBerg \& Bridges 1984; Perryman et
al. 1998) and of field stars (Lebreton et al.  1999) observed by
\emph{Hipparcos}.  In addition, the study of lower main sequence visual binary
systems with known masses and metallicity returns a mixing-length unique and
equal to the solar one for a wide range of ages and metallicities $
\textrm{[Fe/H]}_{\odot} \pm 0.3$~dex (Fernandes et al. 1998).

\begin{figure}
\begin{center}
\includegraphics[scale=0.65]{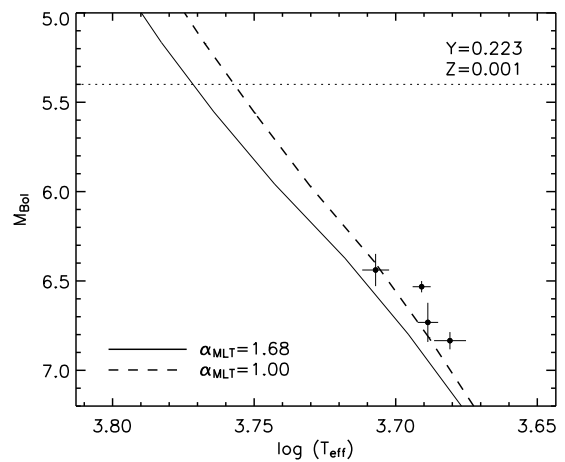} 
\caption{{Effect of changing the mixing length $\alpha_{\rm MLT}$ in a metal
    poor isochrone (age 5 Gyr, although the age does not play any role).  The
    dotted horizontal line is the adopted cut in $M_{Bol}$
    for our sample of stars (see Section \ref{sample}). Overplotted are also
    our dwarfs with $Z \sim 0.001$.}}
\label{mixing}
\end{center}
\end{figure}

Nonetheless, a decrease of the mixing length at low $Z$ would be particularly
interesting since it would produce a less massive convection zone for a given
stellar mass, thus making isochrones cooler.  For metal poor stars this effect
could partly alleviate our ``low helium'' problem. We have tested the effect of
setting the mixing length $\alpha_{\rm MLT}=1.00$ and the difference with
respect to the adopted solar one ($\alpha_{\rm MLT}= 1.68$) is shown in Figure
\ref{mixing} for a moderately helium deficient and metal poor isochrone. As a
result, a large change in the mixing length is indeed able to shift the
isochrone to cooler temperatures, thus improving the agreement with the data.

This result can be regarded as an indication of a metallicity dependence of the
mixing-length for the lower main sequence, as already suggested by Chieffi,
Straniero \& Salaris (1995).  Whether such a large change in $\alpha_{\rm MLT}$
is justified on other evidence or physical grounds remains to be seen.  In this
work we rather test to which extent the most recent low mass stellar models can
be used \emph{ipse facto} to study a large stellar sample in the Solar
Neighbourhood. The range in which models can be safely used is discussed in
Section \ref{bitest}.

\subsubsection{Diffusion}\label{DifSec}

Atomic diffusion (sometimes called microscopic or elemental diffusion) is a
basic transport mechanism which is usually neglected in standard stellar
models. It is driven by pressure, temperature and composition
gradients. Gravity and temperature gradients tend to concentrate the heavier
elements toward the center of the star, while concentration gradients oppose to
the above processes (e.g. Salaris, Groenewegen \& Weiss 2000).  To be
efficient, the medium has to be quiet enough, so that large scale motion cannot
prevent the settling (e.g. Morel \& Baglin 1999; Chaboyer et
al. 2001). Diffusion acts very slowly, with time scales of the order of
$10^{9}$ years so that the only evolutionary phase where diffusion is efficient
is during the Main Sequence\footnote{Diffusion turns out to be important also
in White Dwarf cooling, but this is clearly outside the scope of this paper.},
in particular for metal poor (Population II) stars because of their small
convective envelopes.  For the Sun, the insertion of helium and heavy element
diffusion in the models has significantly improved the agreement between theory
and observations (e.g.  Christensen--Dalsgaard, Proffitt \& Thompson 1993;
Guenther \& Demarque 1997; Bahcall at al. 1997; Basu, Pinsonneault \& Bahcall
2000).  Only in the region immediately below the the convective envelope
theoretical models deviate significantly from the seismic Sun, indicating that
diffusion might not operate exactly in the way calculated or pointing to some
neglected additional physical process partially counteracting diffusion (Brun
et al. 1999).

Due to diffusion the stellar surface metallicity and helium content
progressively decrease during the main sequence phase as these elements sink
below the boundary of the convective envelope. In the deep interior, the
sinking of helium towards the core leads to a faster nuclear aging, thus
reducing the main sequence lifetime with consequences for age determinations of
Globular Clusters (Chaboyer et al. 1992; Castellani et al. 1997).  In the
envelope, diffusion leads to a depletion of the heavy elements and helium thus
producing a decrease of the mean molecular weight.  Metal diffusion decreases
the opacity in the envelope and increase the central CNO abundance: the
dominant effects are the decrease of the mean molecular weights in the envelope
and its increase in the core which increases the model radius and hence
decreases the effective temperature.  The net effect on the evolutionary
tracks, for a given initial chemical composition, is to have a main sequence
cooler. This effect reaches its maximum at the turn-off stage, after which a
large part of the metals and helium diffused toward the center are dredged back
into the convective envelope of giant branch stars, thus restoring the surface
$Z$ and $Y$ to a value almost as high as for evolution without diffusion
(e.g. Salaris et al. 2000).

Diffusion is clearly a major candidate in helping to solve the puzzling low
helium abundances of Section \ref{method}, since it yields a cooler main
sequence, thus operating in the required sense. Besides the effect on the
stellar models themselves, diffusion affects the measured surface metallicities
with respect to the true (original) ones of the stars, altering conclusions
about $Y(Z)$ (see below).  In their pioneering work, Lebreton et al. (1999)
found that main sequence models (for standard values of helium enhancement)
were hotter than \emph{Hipparcos} subdwarfs in the metallicity range $-1 \le
\textrm{[Fe/H]} \le -0.3$. Since decreasing the helium abundance to resolve the
conflict would have required values well below the primordial one (in
accordance to what we have obtained in Section \ref{obstheo} and \ref{method}),
Lebreton et al. (1999) advocated two processes that could help in solving the
discrepancy : \emph{i)} diffusion of helium and heavier elements in stellar
models and \emph{ii)} increase of the measured metallicity in metal poor
objects due to usually neglected NLTE effects.  Correcting isochrones for both
effects they were partly able to solve the discrepancy (see also Morel \&
Baglin 1999), but their number of metal poor and faint stars was rather
modest. Here we test the same corrections on many more stars.

We focus only on the effects of diffusion, leaving the discussion of
observational uncertainties (among which NLTE effects) to Section \ref{nlte}.
The works of Morel \& Baglin (1999) and Salaris et al. (2000) specifically
tackle the effects of helium and heavy elements diffusion in field stars. Both
works assume a full efficiency of the diffusion so that their results can be
regarded as an upper limit on its effects.

From the observational point of view, diffusion decreases the surface
metallicity --provided that it is fully efficient and no other processes
counteract it-- so that a star presently observed with a given [Fe/H] has
started its evolution with a larger metallicity $\textrm{[Fe/H]}_{0}$. As we
discuss later such a difference is of order $0.1$~dex, although sometimes
higher differences have been claimed.  Such a shift in metallicity has
negligible effects on the fundamental parameters of $T_{eff}$ and $M_{Bol}$
determined for our stars with the IRFM (figure 11 in Casagrande et al. 2006),
yet it would imply that our fits in Figures \ref{tehe}--\ref{hema} are
performed with too low Z isochrones, thus needing low $Y$ values to compensate
for the hot isochrone temperatures.

A proper comparison between diffusive and non-diffusive isochrones therefore
must take into account also that diffusive isochrones must start their
evolution with a higher metallicity so that at a chosen age their surface
metallicity (which decreases with time) matches that of non-diffusive
isochrones.  Following the notation of Morel \& Baglin (1999) we call
isochrones that account for both effects (diffusion and correction of the
surface metallicities) ``diffusive calibrated isochrones''.  Diffusion clearly
introduces an age dependence regardless of the fact that stars are still on
their ZAMS. As a general rule, depletion increases with increasing age since
diffusion has more time to work.  Differences between non-diffusive and
diffusive calibrated isochrones are given by Salaris et al. (2000) for various
metallicities, ages and luminosities.  The calibrated diffusive isochrones are
cooler by a few tens up to $100-150$~K, depending on mass and age (see figure 2
in Salaris et al. 2000). However, for the lower main sequence the effect of
diffusion becomes increasingly less significant (their table 1). At the lowest
masses and faintest luminosities covered in this study the effect of diffusion
is at most $30-40$~K in $T_{eff}$.  The reason for such negligible changes is
that in the low mass regime, stars have large convective zones which inhibit
diffusion. Figure \ref{dep} clearly shows that low values of helium are also
found for objects with masses below $0.7 M_{\odot}$, thus suggesting that
diffusion is not the only relevant ingredient to solving our helium
discrepancy.  Similar results to those of Salaris et al. (2000) were also found
by Morel \& Baglin (1999) who give a large set of corrections between non
diffusive and diffusive calibrated isochrones.  Their corrections are provided
for 10~Gyr isochrones in the metallicity range $0.0006 \le Z \le 0.006$ and
masses between $0.6 - 0.85 M_{\odot}$. Their age is chosen in order to maximize
the effect of diffusion. At this age, masses above $0.85 M_{\odot}$ start to
evolve off the main sequence; for masses below $0.6 M_{\odot}$ the effect of
diffusion is negligible.  We apply these corrections to our isochrones and we
consider only masses below $0.85 M_{\odot}$.  We linearly interpolate such
corrections between contiguous values of $M_{bol}$, $T_{eff}$ and $Z$ and apply
them to all our sub-solar metallicities isochrones. In the range $0.006 < Z <
0.017$ we have extrapolated them.  Notice that the corrections in Morel \&
Baglin (1999) are given for isochrones with standard values of $Y(Z)$ whereas
here we apply them to isochrones with a large range of $Y(Z)$. However, the
main effect of diffusion is to alter to surface $Z$ and that does not depend on
$Y$.

\begin{figure}
\begin{center}
\includegraphics[scale=0.4]{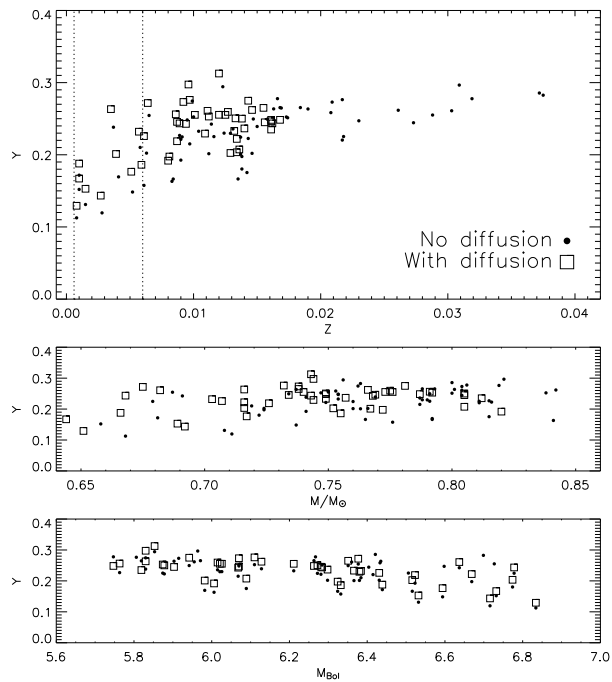} 
\caption{{First panel: $Y$ versus $Z$ plot when the effects of diffusion are 
    included (squares) or not(circles) in the computation. The vertical dotted 
    lines are the metallicity range for which Morel \& Baglin (1999) give 
    corrections. We have extrapolated such corrections up to the solar 
    metallicity $Z=0.017$. For higher metallicities only circles are shown.
    Second and third panel: dependence 
    of $Y$ with Mass and $M_{Bol}$. Error bars are not shown for clarity
    purpose, but are of same magnitude as in Figure \ref{yz} and \ref{dep}.}}
\label{diff}
\end{center}
\end{figure}

The results of computing the helium abundances for our stars with the
corrected isochrones are shown in Figure \ref{diff}. Diffusion clearly helps in
increasing the inferred helium fractions  and its effect --as expected-- 
becomes more important going to lower metallicities. However extremely low 
helium abundances at the lowest $Z$'s are still found.
The fact that low helium abundances are now preferentially found among the
fainter and less massive stars reflects the fact that --as anticipated--
corrections due to diffusion become less and less important descending along
the main sequence. Still, disturbingly low values of $Y \sim 0.2$ remain for
any mass and luminosity, although more orthodox values are within the error
bars.

Until now we have estimated the effect of diffusion in the case of full
efficiency of this process. However there are many observational evidences
suggesting diffusion is less effective.

Diffusion is expected to be more important in metal poor stars, where the mass
of the convective envelope is smaller (e.g. Chaboyer et al. 2001); however,
whether it effectively occurs and how efficiently is still matter of debate,
and especially at low metallicities.  Observations of the narrow Spite
Li--plateau in metal-poor stars (Spite \& Spite 1982; Thorburn 1994; Ryan,
Norris \& Beers 1999; Asplund et al. 2006; Bonifacio et al. 2007) suggest that
diffusion is inhibited near the surface of these objects (e.g.  Deliyannis \&
Demarque 1991; Chaboyer \& Demarque 1994; Ryan et al. 1996) although Salaris \&
Weiss (2001) pointed out that after carefully accounting for uncertainties and
biases in observations, models with diffusion are still in agreement with
observations. More recently Richard et al. (2005) invoked a `turbulent
diffusion' which would limit diffusion without mixing Li. If Li does not allow
firm conclusions, [Fe/H] is a much more robust diagnostic (e.g. Chaboyer et
al. 2001).  The absence of any variation in [Fe/H] between giant branch and
turn-off stars found by Gratton et al. (2001) for two globular clusters
(NGC6397 with [Fe/H]~$=-2.03$ and NGC6752 with [Fe/H]~$=-1.42$) is a very
strong evidence that sedimentation cannot act freely in all stars.  Regarding
field stars, diffusion must affect the measured [Fe/H] only marginally, for
otherwise high-velocity giants in the \emph{Hipparcos} catalogue would have on
average metallicity larger by a factor of two than their turn-off or main
sequence counterparts, a feature which has not been observed (D'Antona et
al. 2005b).

Diffusion changes the slope of the main sequence, rendering it steeper as one
goes to higher luminosities (Morel \& Baglin 1999; Salaris et al. 2000) and
also produces a distortion in the mass-luminosity relation (Morel \& Baglin
1999) so that extremely accurate data could, in principle, detect it.
Interestingly, within the present day accuracy, our results agree with
mass-luminosity relations (see Section \ref{masslum}). Since the efficiency of
diffusion changes with metallicity --if diffusion actually occurs-- a much
larger sample of disk stars than those used in this study (so that the time on
which diffusion has been acting is on average the same and equal to the mean
age of the disk) would probably make possible to detected a change in the slope
of the location of dwarfs with metallicity say, solar and a third of the solar
value.

From the point of view of theoretical modeling, the effect of heavy element
diffusion in metal poor stars is still controversial (e.g.  D'Antona et
al. 2005b; Gratton, Sneden \& Carretta 2004) as theoretical results also differ
according to the formalism employed to describe it.  Models that assume
complete ionization (and then negligible effects of radiation pressure) predict
depletion for all elements heavier than H (e.g. Straniero, Chieffi \& Limongi
1997; Chaboyer et al. 2001).  However, accounting for partial ionization and
radiation pressure shows that whereas some elements like He and Li are expected
to be depleted, others (like Fe) are expected to be significantly enhanced for
stars with $T_{eff} > 6000$~K and only moderately underabundant ($\sim 0.1$~dex
or less) below this temperature (Richard et al. 2002).  Chaboyer et al. (2001)
found that models with full diffusion differ by more than $2 \sigma$ from the
observations of Gratton et al. (2001), thus concluding that heavy-element
diffusion does not occur in the surface layers of metal-poor stars and that
isochrones including the full effects of diffusion should not be used for
comparison with observational data. Although it is not yet clear which
mechanism can counteract diffusion in the surface layers --mass loss (Vauclair
\& Charbonnel 1995), mixing induced by rotation (e.g. Vauclair 1988;
Pinsonneault et al. 1992, 1999, 2002) and radiative diffusion (Morel \&
Th\'evenin 2002) have been proposed among others-- Chaboyer et al. (2001)
found that the temperatures of models in which diffusion is (admittedly ad hoc)
inhibited near the surface (but not in the deep interior) of metal poor stars
are similar to the temperatures of models evolved without diffusion. Also
Richard et al. (2002) concluded that at least in $0.8 M_{\odot}$ stars, it is a
better approximation not to let Fe diffuse than to calculate its gravitational
settling without including the effect of radiative acceleration.

In summary, all models predict the effect of diffusion to increase with
decreasing metallicity, since at lower $Z$ the main sequence shifts to hotter
temperatures, for which convective layers are smaller. At the same time,
Lithium (to some extent) and the most accurate [Fe/H] measurements in globular
clusters (Gratton et al. 2001) pose an upper limit to the effect of diffusion
that even for the most metal poor stars in our sample is expected (if any) to
be negligible or within our error bars. The results shown in Figure \ref{diff}
assume a fully efficient diffusion that is improbable and still do not solve
completely the problem of our low helium abundances.

\subsection{NLTE effects and adopted temperature and luminosity
  scale}\label{nlte} 

As previously noted, Lebreton et al. (1999) were partly able to resolve the low
helium abundance problem by using the cumulated effect of diffusion and NLTE
departures in metallicity measurements.  According to Th\'evenin \& Idiart
(1999) NLTE corrections are negligible for stars with solar metallicity but for
$\textrm{[Fe/H]} \sim -1.0$ the measured metallicity should be increased of
order 0.15 dex (the larger corrections being for hotter --$T_{eff}>6000$~K--
stars that however we do not have in our sample).  Such a difference, although
significant, is roughly of the same order of present day uncertainties in
abundance determinations.  Besides, the relatively large differences claimed by
Th\'evenin \& Idiart (1999) have not been confirmed by other subsequent
studies. Gratton et al. (1999) found negligible departures from LTE in dwarf
stars of any $T_{eff}$ concluding that LTE abundance analysis of metal poor
dwarfs are validated, an important support to the current views on galactic
chemical evolution.  Gratton et al. (1999) also analyzed NLTE effects on
species other than Fe and again they did not find any significant departures in
the case of cool dwarfs. Similar conclusions were drawn by Fulbright (2000) and
Allende Prieto et al. (1999) pointed out that NLTE starts to show up primarily
at $\textrm{[Fe/H]} < -1.0$ (i.e $Z \lesssim 0.003$ whereas our low Y values
are already found at higher metallicities). Thorough calculations accounting
for NLTE effects have been carried out by Gehren et al. (2001a, 2001b), Korn, Shi
\& Gehren (2003) who found negligible corrections for the Sun and up to 0.06~dex
in the case of halo stars. The effects of departures from LTE in abundance
determinations of various elements are widely discussed in Asplund
(2005). Summarizing, in the case of Iron lines a clear consensus about NLTE
effects is still far from reach, but it seems reasonable to assume that
corrections of order 0.10~dex are expected in stars with low metallicities
and/or $\log g$.

The metallicities we use come from various sources (Casagrande et al. 2006), so
that this might account for part of the scatter in the data. However, the
overall trend is clear and therefore does not depend on the specific
metallicity scale adopted.  Besides, in the colour-colour planes (expecially in
the $B-V$ colour index which is very sensitive to metallicity) there is a very
good agreement between our sample of stars and the homogeneous metallicity
scale of model atmospheres (Casagrande et al. 2006).  We have already mentioned
that our work is differential with respect to the Sun and therefore we expect
our results to be unaffected by the new solar abundances obtained when 3D model
atmospheres are adopted, provided that similar updates pertain also to the
lower main sequence stars.  However, large libraries of 3D model atmosphere for
analyzing stellar spectra are not yet available. Therefore, we can not exclude
\emph{a priori} that there are no systematic biases in the models with
$Z$.

We also test whether the low helium abundances depend on the adopted
temperature and luminosity scale. Our empirical IRFM temperature (and
luminosity) scale is in agreement with spectroscopic measurements and $\sim
100$~K hotter than other IRFM temperature scale (see Casagrande et al. 2006 for
a detailed discussion); it closely recovers the temperatures of a set of solar
analogs and indeed the theoretical solar isochrone is in outstanding agreement
with the data (Figure \ref{solar}).  Cooler temperature scales clearly tend to
increase the disagreement with respect to theoretical isochrones, although when
studying the HR diagram it is the combined effect of temperature \emph{and}
luminosity scales which is important. In this respect, the IRFM is one of the
few methods that returns a fully consistent temperature and luminosity scale.

If we adopt the IRFM scale of Ram\'irez \& Mel\'endez (2005) by decreasing our
effective temperatures by 100~K and luminosities by 1.4\% (Casagrande et
al. 2006) the problem of low helium abundances becomes still worse.  The shape
of the $Y$ vs. $Z$ plot is the same (reflecting the offset in the absolute
calibration adopted, Casagrande et al. 2006) but the helium content is on
average lower by $\sim 0.04$ so that already at solar metallicity the bulk of
stars has a helium content lower than our solar calibrated model.

\section{The binary test}\label{bitest}

To date, the most stringent tests of the theory of stellar structure and
evolution have been carried out for the Sun. Its mass, luminosity and radius
are known to better than 1 part in $10^3$ and its age to better than few
percent (e.g. Guenther \& Demarque 2000; Bahcall, Serenelli \& Basu 2006).
Its chemical abundance, which sets the zero point of metallicity measurements
in other stars, is currently under profound discussion (Asplund et al. 2005)
however, as we have already mentioned, this change should not affect
dramatically our study since our work is differential with respect to the Sun.
The Sun is therefore the natural benchmark in understanding and setting models
of stellar structure and evolution. We have already checked the solar isochrone
to be in excellent agreement with our solar metallicity stars (Figure
\ref{solar}).

In the case of stars other than the Sun, radii and luminosities are known with
much less accuracy although interferometry is expected to be a major
breakthrough in the next few years. For the moment, masses can be empirically
determined only in the case of systems in binaries. For visual binaries with
well-measured parallaxes, the uncertainty in mass determination is rarely less
than 1\%, a value that sets the accuracy required to provide important
constraints on models of stellar structure and evolution (e.g. Andersen 1991).
In addition to the mass, the measured colours and metallicities are another
source of errors.

As we have already discussed in Section \ref{method} such limitations preclude
the accurate calibration of stellar models on binary stars. Our model is
calibrated on the Sun but the comparison with a statistically congruous number
of binaries can indeed provide important constraints on it.  Here we use
various double stars with accurately measured masses, metallicities and colour
indices for at least one of the components. They are all nearby, so that no
reddening corrections are needed. Colours and metallicities are used to derive
$T_{eff}$ and $M_{Bol}$ consistently with our IRFM scale. The mean metallicity
from various recent measurements is used so as to reduce the uncertainty in
this observable. The same procedure as described in Section \ref{method} is
then applied to deduce the mass and the helium content of these binaries.
Although the broadening (and so the helium content) of the lower main sequence
is independent of assumptions about stellar age, masses are not, as we have
quantified in Section \ref{method}.  Here we are interested in testing to what
extent our choice of using 5~Gyr old isochrones is on average able to recover
the masses of lower main sequence dwarfs. Though with a large scatter such age
should be in fact representative of the Solar Neighbourhood (Reid et al. 2007),
also considering there is no clear consensus on the tightness of the
age--metallicity relation (e.g. Feltzing et al. 2001), so that older isochrones
are not necessarily the most appropriate for metal poor stars.  The masses
deduced from the isochrones are compared to those measured empirically: if the
masses are recovered with an accuracy of few percent the corresponding helium
content of the stars --which is practically age independent-- is also 
validated (Figure \ref{hema}).

Note that to have a congruous number of stars, we have slightly relaxed our
cutoff on $M_{Bol}$. Possible evolutionary effects have therefore been taken
into account for the brightest stars, but if not otherwise specified the age
adopted for the isochrones is fixed to 5~Gyr.  We also mention that all these
stars belong to non-interacting binary systems so they are representative of
single stars.

\subsection{$\alpha$~Cen B}

Among stars other than the Sun, the $\alpha$~Cen system is probably the most
used test-bed for checking stellar models (see also discussion in Section
\ref{mixi}).  Its secondary component (HD 128621) is a K dwarf and it has been
a privileged target for asteroseismic (e.g. Th\'evenin et al. 2002; Kjeldsen et
al. 2005) and interferometric (Kervella et al. 2004; Bigot et al. 2006)
studies. Since it is in a well separated binary this K dwarf is part of our
original sample of Section \ref{sample}, but here we analyze the results in
more detail.  Using positions and radial velocities, its mass has been
estimated to great accuracy ($M= 0.934 \pm 0.0061\,M_{\odot}$) and completely
independently of theoretical considerations of stellar structure and evolution
(Pourbaix et al. 2002).  For this star there are various independent and
accurate metallicity measurements (Valenti \& Fischer 2005; Santos et al.
2005; Allende Prieto et al. 2004; Feltzing \& Gonzalez 2001) with a mean value
$\textrm{[Fe/H]} = 0.23 \pm 0.03$~dex and solar scaled abundances.  Accurate
$BV(RI)_C$ colours (Table 3) are available from Bessell (1990) from which
$T_{eff}$ and $M_{Bol}$ are computed as described in Section \ref{sample}.  We
obtain a mass of $0.925 \pm 0.035 M_{\odot}$ in excellent agreement with that
measured empirically. The corresponding helium content is found to be $0.262
\pm 0.022$ and therefore equal to the solar one within errors, although the
star is more metal rich.

\subsection{vB22}

As summarized by Lebreton et al. (2001), the Hyades cluster has five binaries
whose components have measured masses. Of these systems only the eclipsing
binary HD 27130 (vB22) has masses with small enough uncertainty to place
significant constraints on theoretical models, as studied by Pinsonneault et
al. (2003).  $BV(RI)_C$ magnitudes and colours of both components are available
from Schiller \& Milone (1987) and are listed in Table 3. Again, $T_{eff}$ and
$M_{Bol}$ are derived according to the procedure described in Section
\ref{sample}.  Though metallicity measurements for an eclipsing binary are
quite uncertain, we exploit the fact that the metallicity of such a system must
be the same of other cluster members. For the Hyades, Paulson, Sneden \&
Cochran (2003) have conducted a detailed spectroscopic analysis from which a
mean metallicity $\textrm{[Fe/H]} = +0.13 \pm 0.04$~dex and solar scaled
abundances have been derived. High-precision distance estimates are available
from \emph{Hipparcos} ($\omega=21.40 \pm 1.24$) and from a kinematic parallax
(de Bruijne et al. 2001, $\omega = 21.16 \pm 0.38$~mas). These are all in
excellent agreement and we assume de Bruijne et al. (2001) measurement in the
following.

Empirical masses are available from Torres \& Ribas (2002).  For the mass and
bolometric magnitude of the primary, age effects become relevant and, rather
than our standard reference 5~Gyr isochrones, we consider isochrones of 500~Myr
(1~Gyr), consistent with the age of the cluster (Perryman et al. 1998). The
estimated mass is 1.100 (1.088) $M_{\odot}$ ($1 \sigma$ more massive than the
empirical value) and the corresponding helium content $Y \sim 0.22$. Optimizing
on the mass formally returns an age of 2.48~Gyr and $Y \sim 0.23$. Evidently
age effects are important here, but in any case, the helium content of the
system is significantly below solar.  For the secondary, as expected, the
helium content is independent of the age chosen for the isochrones. The
difference in the use of 500~Myr and 5~Gyr isochrones is less than 0.02 in mass
and 0.004 in helium abundance, i.e. smaller than the uncertainty of the
results.  Using 5~Gyr isochrones, the mass we recover for the secondary is in
very good agreement with the empirical one, with a helium content significantly
lower than the solar one.  Though depending also on the exact metallicity of
the binary (we have assumed the average value of the cluster, but a slight
scatter among its stars is possible), our result provide a further, strong
evidence that the Hyades are underabundant in helium for their metallicity
(Perryman et al.\ 1998; Lebreton et al.\ 2001; Pinsonneault et al.\ 2003).

\subsection{$70$~Oph}

$70$~Oph (HD 165341) is one of our nearest neighbours and is among the first
discovered binary stars. Gliese \& Jahrei\ss~(1991) classify it as a primary of
spectral type K0 V and a secondary of type K5 V. Recent abundance analysis for
the primary companion are available from Luck \& Heiter (2006), Mishenina et
al. (2004), Allende Prieto et al. (2004). We adopt the mean value
$\textrm{[Fe/H]} = -0.02 \pm 0.08$~dex and $\textrm{[} \alpha \textrm{/Fe]} =
0.06 \pm 0.11$~dex.  Tycho $B_T$ and $V_T$ magnitudes for both components are
available from Fabricius \& Makarov (2000) which we convert to the
Johnson-Cousins system by interpolating the transformation coefficients given
in table 2 of Bessell (2000). Additional photometry is available from Gliese \&
Jahrei\ss~: $V$ magnitudes agree with the transformed Fabricius \& Makarov
(2000) ones within $0.01$~mag, whereas the $B-V$ index of Gliese are slightly
($0.02-0.04$~mag) redder. $(R-I)_K$ from Gliese \& Jahrei\ss~has been converted
to the Cousins system with the transformation given in Bessell (1995). For
consistency with the choice made for most of the binaries we use only
magnitudes and colours from Gliese \& Jahrei\ss. Averaging with the $B$ and $V$
magnitudes from Fabricius \& Makarov (2000) hardly changes the results : the
large uncertainty in $T_{eff}$ for the primary component remains the same and
the changes in the helium content and masses of both components are smaller
than their final errors.  We use the \emph{Hipparcos} parallax and errors as
corrected by S\"oderhjelm (1999) ($\omega = 195.70 \pm 0.90$~mas) for binarity
effects. Masses for the primary and the secondary are available from Henry \&
McCarthy (1993) ($M=0.856 \pm 0.056\,M_{\odot}$ and $M=0.713 \pm
0.029\,M_{\odot}$, the secondary as recomputed with an improved parallax by
Delfosse et al. 2000) and from Fernandes et al. (1998) ($M=0.89 \pm
0.04\,M_{\odot}$ and $M=0.71 \pm 0.04\,M_{\odot}$) that we assume in the
following\footnote{Note that Fernandes et al. (1998) calibrated stellar models
on some of the binaries we also discuss in this Section. Their approach is
quite different from ours since they had helium content, age, mixing-length and
individual masses of both components as free parameters in the model. However,
they also computed empirical masses of both components and used the total mass
as a constraint on the model. In the following Section we only use their
empirical masses for comparing our results.}. Based on asteroseismic
considerations Carrier \& Eggenberger (2006) also derived a mass of
$0.87\,M_{\odot}$ for the primary component. The masses we derive are in good
agreement with the empirical ones although the large uncertainty in $T_{eff}$
of the primary returns considerably large error bars.  The helium content is
equal to the solar one within the errors, consistently with the expectation
given the solar metallicity. A solar helium abundance was also obtained by
Fernandes et al. (1998).

\subsection{HD 195987}

Combining spectroscopic and interferometric observations for this double-lined
binary system, Torres et al. (2002) derived masses with a relative accuracy of
a few percent. They also determined the metallicity ([Fe/H]$=-0.5$,
[$\alpha$/Fe]$=+0.4$ and uncertainty $\sim 0.2$~dex), orbital parallax
($\omega= 46.08 \pm 0.27$~mas in rough agreement with the \emph{Hipparcos}
value, but with smaller formal error) and $V,H,K$ magnitudes for both
components. Their infrared magnitudes are in the CIT system and we convert them
into the 2MASS by using the Carpenter (2001) transformations. We then use our
effective temperature and bolometric luminosity calibrations and the procedure
described in Section \ref{method} to deduce the mass and helium content of both
components. The mass of the primary is higher by $3 \sigma$ but that of the
secondary is in good agreement with the empirical value.  Both components are
fitted with similar (and well below primordial) helium content. Ascribing the
mass discrepancies to temperature effects, an increase of 70~K in the $T_{eff}$
of the secondary (or more properly a corresponding decrease in the effective
temperature of the isochrones) would return a mass ($0.666 M_{\odot}$) in
excellent agreement with the empirical one, but the helium content would be
still very low ($Y=0.180$). For the primary the temperature should be increased
by 300~K in order to obtain a mass ($0.842 M_{\odot}$) in agreement with the
empirical one. In this case the helium content would be $Y=0.269$, higher than
that of our solar calibrated model. However, the primary is very luminous given
its mass, so that it could be a slightly evolved stars and therefore 5~Gyr
isochrones could not be the most appropriate choice. If 10~Gyr old isochrones
are used, masses are decreased so that the primary is off by $2 \sigma$ and
both components are again fitted with similar helium content.

\subsection{$\xi$~Boo}\label{xiBo}

$\xi$~Boo (HD 131156) consists of a primary of spectral type G8 V and a
secondary K4 V (Gliese \& Jahrei\ss~1991). The primary is known to be very
active, with irregular fluctuations of activity (e.g. Petit et al. 2005 and
references therein) and a high chromospheric emission (Baliunas et al. 1995)
being classified as flare star in SIMBAD and variable in \emph{Hipparcos}.
These data suggest a young age that also agrees with conclusions from
evolutionary models (Fernandes at al. 1998).  Recent abundance analyses for the
primary component are available from Luck \& Heiter (2006), Valenti \& Fischer
(2005), Allende Prieto et al. (2004), Fuhrmann (2004). We adopt the resulting
mean value $\textrm{[Fe/H]} = -0.15 \pm 0.09$~dex and $\textrm{[} \alpha
\textrm{/Fe]} = -0.06 \pm 0.15$~dex.  $V$ magnitudes, $B-V$ and $(R-I)_K$
indices of both components are available from Gliese \& Jahrei\ss~(1991). We
convert $(R-I)_K$ into Cousins system by means of the Bessell (1995)
transformations.  Improved \emph{Hipparcos} parallaxes are available from
S\"oderhjelm (1999) and empirical masses from Fernandes et al. (1998).

The results are certainly interesting : the primary is more massive than the
empirical value, whereas the secondary is $2 \sigma$ less massive. Also, the
helium content between the two component differs by $1 \sigma$, whereas all the
other binaries in this study have identical helium abundances within the
errors. To reach a closer agreement with the empirical masses, that of the
primary should be decreased (implying a higher helium abundance, see Figure
\ref{hema}) whereas that of the secondary should be increased (thus lowering
its helium content). Of course, the photometry could be adversely affected by
the young system age and be the simple explanation for these puzzling results.
Torres et al. (2006) found chromospheric activity as a likely cause of the
discrepancy between models and observations in the case of another star (HD
235444). We have decided to discard this binary from our basic sample of stars
with empirical masses.

One might still wonder whether variability can occur among some of the stars in
Section \ref{sample} and to what extent this might be behind our anomalous
helium abundances.  Extensive surveys by \emph{Einstein} and \emph{ROSAT} and
\emph{Chandra} X-ray satellites have shown that late-type main sequence stars
are surrounded by coronae analogous to the more easily observed solar corona
(e.g. Schmitt \& Liefke 2004; Wood \& Linsky 2006).  Flares, spots, coronal
mass ejections, prominences are, of course, not exclusive to our Sun. For our
sample of stars we have used the $\Delta \mu$ method to remove unresolved
binaries (whose tidal interaction could trigger activity, e.g. Torres et
al. 2006) and the same sample is also free from variable stars to a high
accuracy level (Casagrande et al. 2006).  Therefore, any intrinsic level of
activity in the sample of Section \ref{sample} is below our observational
uncertainties and is unlikely to be causing our helium
discrepancies. Furthermore, to explain the low helium abundances, variability
should practically be limited to the metal poor stars, whereas variability is
known to occur at all metallicities.

\subsection{$\eta$~Cas~B}

$\eta$~Cas (HD 4614) is a nearby visual binary at a distance $\sim 6$~pc.
According to Gliese \& Jahrei\ss~(1991) it consists of a primary of spectral
type G3 V and a secondary K7 V. $\eta$~Cas~A is known to be over-luminous with
respect to the mass-luminosity relation, thus suggesting that it has begun to
evolve off the main sequence (Fernandes et al. 1998).  In what follows we focus
on the secondary, $\eta$~Cas~B. Abundance analyses for cool dwarfs are still
challenging (e.g. Bonfils et al. 2005), however the metallicity of the primary
is well determined. We have taken five independent metallicity measurements
(most of which include $\alpha$-elements) from Luck \& Heiter (2006), Valenti
\& Fischer (2005), Bonfils et al. (2005), Mishenina et al. (2004), Allende
Prieto et al. (2004). All these measurements show good agreement and we adopt a
mean value of $\textrm{[Fe/H]} = -0.31 \pm 0.07$~dex and $\textrm{[} \alpha
\textrm{/Fe]} = 0.10 \pm 0.03$~dex. Visual magnitude, $(B-V)$ and $(R-I)_K$
colours for the secondary are also available from Gliese \&
Jahrei\ss~(1991). $(R-I)_K$ is in the Kron system and it has been converted to
the Cousins system with the transformation given in Bessell (1995).  Both
colours are slightly redder than the applicability range of our temperature and
bolometric luminosity calibrations, consistently so with the late spectral type
of this star. However the mean loci of the calibrations in Casagrande et
al. (2006) (see their figure 13 and 18) show well defined trends so that the
extrapolation to stars $\sim 100$~K cooler than the applicability range is
still quite reasonable.  The parallax of the primary is available from
\emph{Hipparcos} and the mass of the secondary from Fernandes et al. (1998).
Notice that both the spectral type and the mass of this star are slightly lower
than that of our sample stars in Section \ref{sample}. The result for this
moderately metal deficient star is particularly interesting, since it does
yield a low helium abundance (but in agreement with the bulk of stars with the
same metallicity in Figure \ref{yz}) while its derived isochrone mass is lower
than the empirical value. At this metallicity NLTE effects in the derived
metallicities should be very small if any, as well as the effect of diffusion
that are negligible at $0.6 M_{\odot}$.  Also, from Figure \ref{hema} an
increase of the mass at fixed metallicity would require an even lower helium
abundance.

As noted above, Fernandes et al. (1998) calibrated stellar models to this
binary, obtaining a helium content ($Y=0.25$) higher than that found
here. However, considering the difference in the respective solar reference
value (their solar model has $Y_{\odot}=0.28$) our result is within $1 \sigma$
with theirs.

\subsection{$85$~Peg~A}

85 Peg (HD 224930) is a well studied, metal poor, visual and single-lined
spectroscopic binary. Its small angular separation and the marked magnitude
difference between the components ($\Delta m_V=3.08 \pm 0.29$~mag, ten
Brummelaar et al. 2000), makes it a difficult target both for visual and
spectroscopic observations. The given $\Delta m_V$ implies a magnitude
correction of $0.06$~mag for the primary that increases going to longer
wavelength ($0.12$~mag in $I$, see equation \ref{deltam}).  The contribution of
the secondary thus must be properly removed.  The total mass of the system is
well constrained from visual orbital elements and Kepler's third law
(e.g. Griffin 2004; Fernandes et al. 2002) however the masses of the individual
components are much more uncertain.

For the primary 85 Peg A, we adopt a mass of $0.84 \pm 0.08 M_{\odot}$, as
obtained by Fernandes et al. (2002) and in near agreement with D'Antona et
al. (2005b) who, based on model predictions, estimated the mass to be in the
range 0.75 to 0.82 $M_{\odot}$. For many years investigators have claimed that
85 Peg B, the fainter companion, is more massive than 85 Peg A, the brighter
one (e.g. Hall 1948; Underhill 1963; Heintz 1993) and this abnormal situation
could be explained if 85 Peg B is an undetected binary, as already suggested by
Hall (1948). Indeed, this seems to be the case, as the same conclusions can be
drawn when the position of both components in the HR diagram is compared with
theoretical expectations (Fernandes et al. 2002; D'Antona et al. 2005b). In
what follows we only consider the primary component.  We use the
\emph{Hipparcos} parallax and errors as corrected by S\"oderhjelm (1999)
($\omega = 82.50 \pm 0.80$~mas) for binarity effects.  For the metallicity we
adopt the mean value obtained from six recent independent determinations (Luck
\& Heiter 2006; Mishenina et al. 2004; Allende Prieto et al. 2004; Fuhrmann
2004; Gratton et al. 2003; Fulbright 2000). All these measurements show good
agreement with an average $\textrm{[Fe/H]} = -0.90 \pm 0.06$~dex and
$\textrm{[} \alpha \textrm{/Fe]} = +0.40 \pm 0.04$~dex.  Spectroscopic
determinations return temperatures ranging from $\sim 5300$~K (Fulbright 2000)
to $\sim 5600$~K (Fuhrmann 2004), the latter being in close agreement with
other recent determinations obtained by carefully fitting Balmer line wings
(D'Antona et al. 2005b).  A temperature of about $5600$~K seems to be favored
also by stellar modeling (Fernandes et al. 2002; D'Antona et al. 2005b).  By
means of adaptive optics, ten Brummelaar et al. (2000) obtained $VRI$
differential photometry of the components, from which individual magnitudes
were then deduced using the composite magnitudes found in the General Catalogue
of Photometric Data (Mermilliod, Mermilliod \& Hauck 1997). With the individual
magnitudes given by ten Brummelaar et al. (2000), we obtain $T_{eff}=5000 \pm
350$~K, $M_{Bol}=5.06 \pm 0.07$~mag and an angular diameter that, translated
into linear radius via parallax, of $1.15 \pm 0.15 \, R_{\odot}$.  Besides a
temperature much smaller than other determinations, the returned linear radius
seems to be excessively large for a star with a mass well below that of the
Sun. Furthermore, in the Catalogue of Absolute Radii of Stars compiled by
Pasinetti Fracassini et al. (2001) the mean linear radius obtained with various
indirect techniques is $0.84 \, R_{\odot}$.  Prompted by such a puzzling
result, we have made an independent search for accurate photometry, rather then
using a generic mean value. In particular, we caution that Johnson $RI$ bands
lack a clearly defined set of standard (e.g. Bessell 1979; Fernie 1983) in
contrast with the excellent $(RI)_C$ system defined by Cousins.  Accurate
photometry is available from Eggen (1979) who observed this star in the
Eggen-Kron $(RI)_K$ system defined by Eggen (1968, 1975).  An excellent
representation of this system is available from Weis (1983, 1996) for which
accurate transformations to the Johnson-Cousins system are given by Bessell
(1995). Adopting this transformation, we obtain the following composite
magnitudes ($V=5.75, V-R_C=0.34, (R-I)_C=0.42$) from which the magnitudes of
the primary in different colours ($m_A$) can be calculated using the following
equation:\footnote{As a further consistency check of our procedure, we point
out that the composite $(V-I)=0.96$ given by ten Brummelaar et al. (2000) can
be transformed to the Cousins system with the recent calibration given by An et
al. (2007) in their appendix A. Such a transformation returns $V-I_C=0.76$,
identical to what we obtain.}
\begin{equation} \label{deltam}
m_A = m + 2.5 \log(1+10^{-0.4\,\Delta m}),
\end{equation}
where $\Delta m$ is the differential photometry in the given colour.  By doing
so we obtain the magnitudes and colours of the primary listed in Table 3 from
which $T_{eff}=5730 \pm 340$~K, $M_{Bol}=5.27 \pm 0.05$~mag and $R = 0.80 \pm
0.09 \, R_{\odot}$. The effective temperature and linear radius are now in much
better agreement with other determinations.  The striking difference of these
parameters with respect to the previous values (though the errors are still
similar) is likely to be due to the shift in the zero-points when the
appropriate photometric system is adopted. The errors are still large, but that
is mostly due to the uncertainties in $\Delta m$ that are of the order of
$0.3$~mag.

For this star we obtain a mass practically identical to the empirical measured
one and a rather low helium content, although the errors in both empirical and
theoretical mass determinations are unfortunately very large. Interestingly
Fernandes et al. (1998) could not calibrate stellar models to this system
unless they assumed extremely low helium abundance $Y<0.2$ and high age
($>20$~Gyr) both at odds with cosmological constraints.  In our case this
cosmological problem is alleviated, but not solved. Fernandes et al.  (2002)
were finally able to calibrate a stellar model including diffusion with a
reasonable age (9.3 Gyr) and helium content ($Y=0.253$), however it was only by
assuming an initial metallicity $\textrm{[Fe/H]}=-0.185 \pm 0.054$ i.e. almost
0.4~dex higher than their adopted observed metallicity ($\textrm{[Fe/H]}=-0.57
\pm 0.11$). Such a difference of $\sim 0.4$~dex would imply a huge effect due
to diffusion, and is supported neither by observations (see discussion in
Section \ref{notguilty}) nor by theoretical modeling: for this object D'Antona
et al. (2005b), assuming an age of 12~Gyr such as to maximize the effect, found
a difference of 0.12~dex due to metal diffusion. Furthermore, their adopted
observed metallicity is 0.3~dex higher than the most recent determinations
([Fe/H]=--0.90, see above), a difference that not even NLTE can easily explain.

\subsection{$\mu$~Cas~A}

Another interesting metal poor system is the halo binary $\mu$ Cas (HD 6582)
which, from the pioneering work of Dennis (1965), has been extensively studied
for its potential role in determining the primordial helium abundance
(e.g. Catchpole, Pagel \& Powell 1967; Hegyi \& Curott 1970; Haywood, Hegyi \&
Gudehus 1992). These older works had to deal with much less accurate estimates
of mass, luminosity, temperature, metallicity and pre-\emph{Hipparcos} parallax
so that they estimated helium abundances ranging from 0 (Hegyi \& Curott 1970)
to 0.4 (Catchpole et al. 1967) with a preference around 0.2 (Haywood et
al. 1992).  While the high magnitude difference between the two components in
the optical ($\Delta m_v=5.5$, McCarthy et al. 1993) has for a long time
hindered accurate relative angular separation measurements and therefore
precise mass determinations, the huge luminosity difference makes almost
negligible the contribution of the secondary to optical photometry. According
to McCarthy et al. (1993), $\Delta m=5.5 \pm 0.7$ at $0.55\;\mu$m (roughly $V$
band) and decreases to $\Delta m=4.5 \pm 1.0$ at $0.75\;\mu$m (roughly $I$
band). Using equation (\ref{deltam}) we can estimate the contribution of the
secondary in $V$ band to be only 0.007~mag, whereas increases to 0.017 mag in
$I$ band.  We only study the primary component.  We have taken the mean
composite Johnson $BV$ magnitudes found in the General Catalogue of Photometric
Data (Mermilliod, Mermilliod \& Hauck 1997) and $RI_K$ magnitudes from Eggen
(1973) then converted to the standard Cousins system with the transformation
given in Bessell (1995).  $V(RI)_C$ magnitudes have then been corrected
according to equation (\ref{deltam}) to account for the contribution of the
secondary.  Although the corrections are at the same level of the photometric
accuracy, they avoid the introduction of systematics in the zero-point. The
magnitude difference between the two components in $B$ band is not available,
but in this band the contribution of the cool secondary component is certainly
below a few millimag.  We have used the \emph{Hipparcos} parallax and for the
metallicity we have taken the mean value from six recent determinations (Luck
\& Heiter 2006; Mishenina et al. 2004; Allende Prieto et al. 2004; Fuhrmann
2004; Gratton et al. 2003; Fulbright 2000). All determinations agree remarkably
well, with $\textrm{[Fe/H]} = -0.91 \pm 0.05$~dex and $\textrm{[} \alpha
\textrm{/Fe]} = 0.36 \pm 0.04$~dex.  The determination of the empirical masses
of both components has been troublesome because of the aforementioned
luminosity difference, however the most recent data agree and we use the
Drummond, Christou \& Fugate (1995) mass of the primary $0.742 \pm 0.059
M_{\odot}$ that becomes $0.757 \pm 0.059 M_{\odot}$ after accounting for the
better parallax provided by \emph{Hipparcos} (Lebreton et al. 1999).

We obtain a mass a little more than $1 \sigma$ higher than the empirical
value. Again, we keep temperature as a free parameter to investigate by how
much it should change to exactly reproduce the observed mass. In this case an
increase of 175~K in the effective temperature would be enough to reduce the
mass to the observed value and to increase the helium abundance to $Y=0.257$.
Th\'evenin \& Idiart (1999) found a correction of +0.15~dex for the metallicity
of this star because of NLTE effects. An increase of the metallicity by such an
amount changes its mass to $0.823 M_{\odot}$ and $Y=0.215$, but to exactly
recover the measured mass, $T_{eff}$ should be still increased by 145~K,
implying a helium abundance $Y=0.273$, higher than the solar value.

\subsection{What we learn from binaries}

\begin{figure*}
\begin{center}
\includegraphics[scale=0.70]{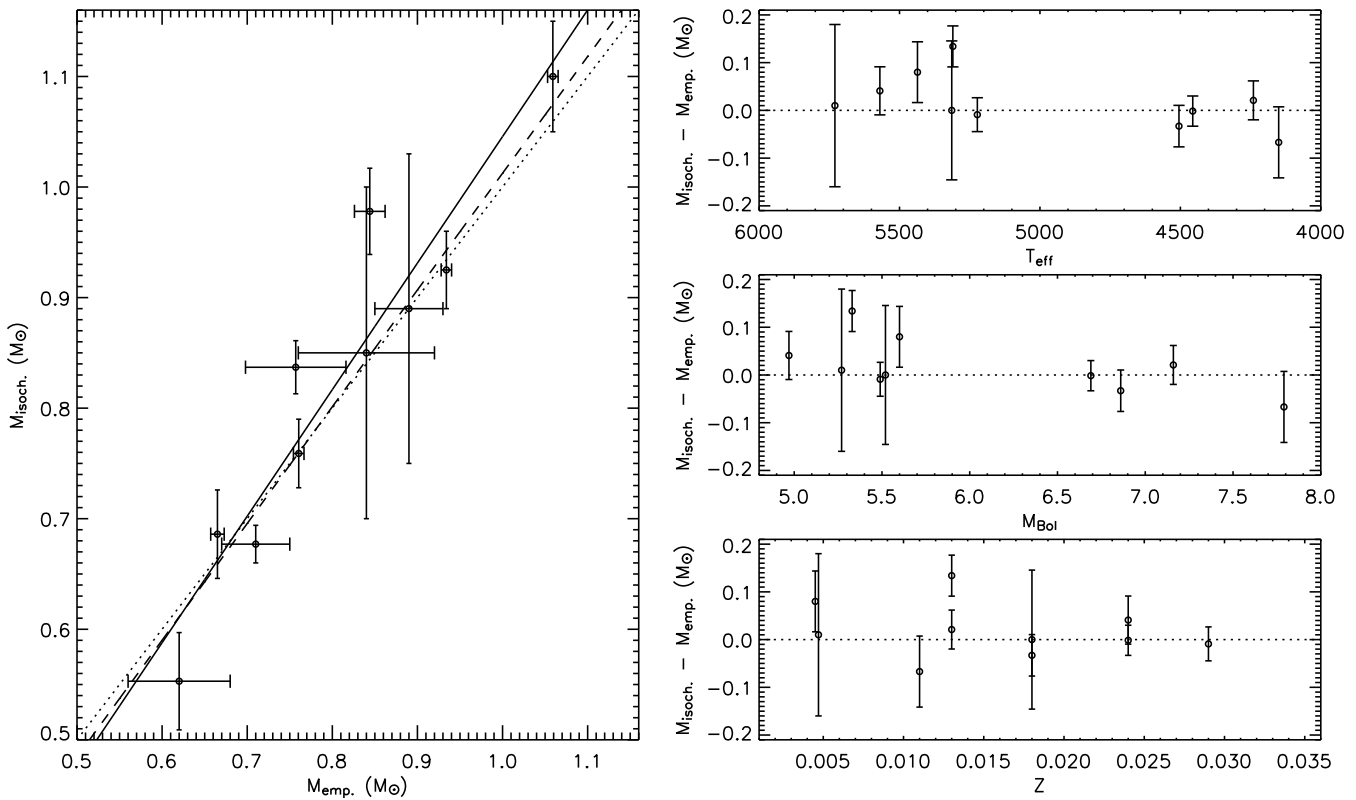}
\caption{{Empirical vs isochrones' masses for the binaries studied in Section
    \ref{bitest}. Dotted lines are the 1--to--1 relation intended to guide the
    eye. The continuous line is the fit of the empirical vs. isochrones' 
    masses; the dashed line is the same fit, when the two possibly evolved 
    stars vB22~A and HD 195987~A are excluded.}}
\label{fitmass}
\end{center}
\end{figure*}

The comparison between empirical and isochrone derived masses for the sample of
binary stars is shown in Figure \ref{fitmass}. Note that neither component of
$\xi$~Boo is included in the comparison, due to its high variability.

The overall agreement is very good and there is only one star (HD 195987~A)
that deviates by $3 \sigma$. Also, the fit pivots around $0.6 - 0.7 M_{\odot}$
and below $\sim 0.9 M_{\odot}$ it is close to the one-to-one relation, with a
systematic difference of at most $0.03 - 0.04 M_{\odot}$. It is possible for
the fit to diverge from the one-to-one relation when going to higher masses
because of possible evolutionary effect associated with vB22~A and HD 195987~A,
as we have already discussed. If these two objects are neglected, the fit is in
outstanding agreement with the one-to-one relation (Figure \ref{fitmass}).

When the difference between empirical and isochrone determined masses is shown
as function of metallicity, it appears that the most serious discrepancies
arise at low metallicity. These differences have been extensively discussed on
a case by case basis above: the simplest way to achieve concordance is to
change in the measured metallicities or a adopt a cooling of the isochrones or
both, but in all cases the changes needed are significantly beyond our expected
uncertainties in these parameters.

Also, it is clear from Figure \ref{fitmass} that at low metallicities
isochrones preferentially tend to overestimate masses; for a given age, a
decrease of the masses would actually imply a higher helium content (Figure
\ref{hema}).  The use of older isochrones at the lowest metallicities could
indeed reduce the discrepancy but would only improve the agreement in mass,
leaving mostly unchanged the anomalously low helium abundances in these stars.

A thorough investigation would require us to simultaneously change temperature,
luminosity and metallicity to find the set of solutions that better fit the
empirical masses. However, our aim here is more modest, to check over what
effective temperature and metallicity range our isochrones return masses in
agreement with the empirical masses. It turns out that for metallicities above
solar the agreement is always within $1 \sigma$ and this strengthens the
conclusion we already reached in Section \ref{method}. At low metallicity, in
spite of the large uncertainties of the individual determinations, the
occurrence of very low $Y$ values, well below the primordial level, is also
confirmed. Our conclusion is that, most likely, there is as yet untreated
physics in lower main sequence, metal poor stellar models.

\begin{figure*}
\begin{center}
\includegraphics[scale=0.82,angle=180]{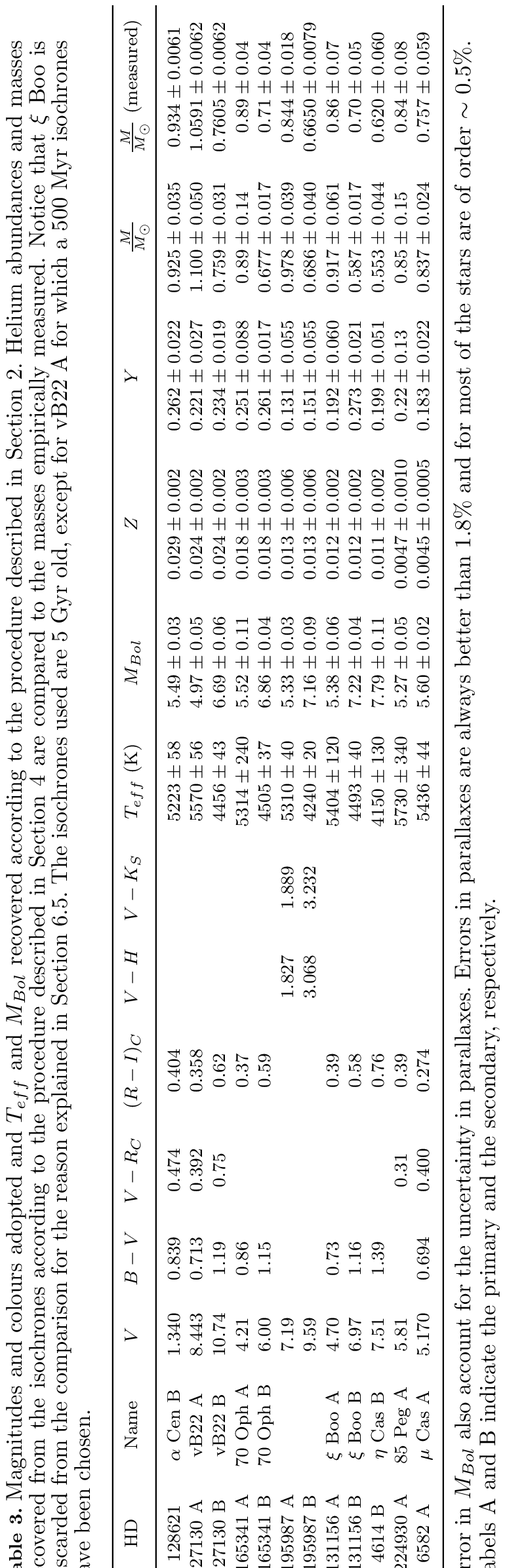} 
\end{center}
\end{figure*}

\section{Mass-Luminosity relation}\label{masslum}

\begin{figure*}
\begin{center}
\includegraphics[scale=0.70]{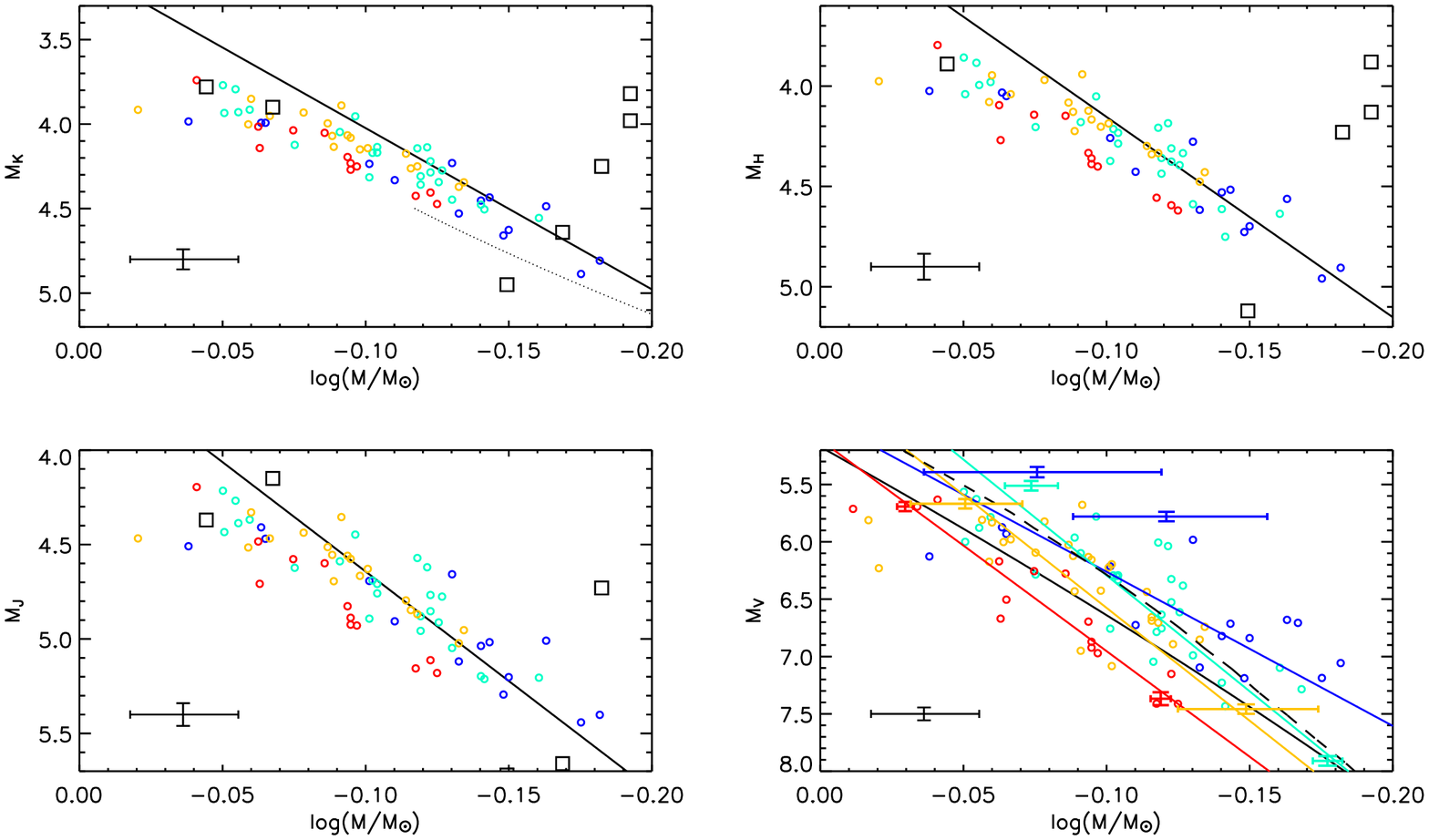}
\caption{{Empirical mass-luminosity relation from Henry \& McCarthy (1993) in
    different bands (solid black line) overplotted to our sample stars.  The
    squares are the stars used by Henry \& McCarthy (1993) to fit their
    empirical mass-luminosity relation.  Infrared colours have been converted
    to the CIT system. Only stars with accurate IR photometry
    (``j\_''$+$``h\_''$+$``k\_msigcom''$< 0.10$) are shown. Points correspond
    to the sample stars in the range $Z < 0.007$ (blue), $0.007 \le Z < 0.014$
    (cyan), $0.014 \le Z < 0.022$ (yellow), $Z \ge 0.022$ (red). The dotted
    line in the first panel is the Delfosse et al. (2000) empirical
    relation. The dashed line in the fourth panel is the Kroupa et al. (1993)
    empirical relation. In the fourth panel are also shown (with colored error
    bars) the stars of Table 3 with the exception of $\xi$~Boo A and B (see
    discussion in Section \ref{bitest}), vB22~A and $\eta$~Cas~B (outside of
    the plot range). The coloured lines are fits in the same
    metallicity bins of the sample stars as explained in the text. 
    A typical error bar for the points is also shown
    in the lower left of each panel.}}
\label{malu}
\end{center}
\end{figure*}

In the previous Section we have compared the masses deduced from the isochrones
to those directly measured for a congruous set of binary stars. Another
approach is to compare empirical mass-luminosity relations to the theoretical
masses (i.e. deduced from the isochrones) and the empirical luminosities
available for our 86 stars of Section \ref{sample} (Figure \ref{malu}).

For the comparison we use the empirical mass-luminosity relation of Henry \&
McCarthy (1993) which extends from $1 M_{\odot}$ down to $0.08
M_{\odot}$. Recent improvements to this relation (e.g. Henry et al. 1999;
Delfosse et al. 2000) concern only the very low mass regime and is not
applicable to our K dwarfs. Henry \& McCarthy's (1993) relation is given in the
infrared ($J,H,K$) and visible ($V$) bands. Since most of the infrared
photometry used by Henry \& McCarthy (1993) is in the CIT system, we have
converted our 2MASS colours via transformations in Carpenter (2001). The
corrections are typically of a few 0.01 magnitudes, whereas the larger
uncertainties in the empirical relations actually come from the 0.03-0.06
scatter in $\log(M/M_{\odot})$.  Another empirical mass-luminosity relation in
$V$ band is that of Kroupa, Tout \& Gilmore (1993) who followed a different
approach to derive it. Rather than fit the data directly, Kroupa et al. (1993)
adopted a reference luminosity function, and chose a functional form for the
mass function. The mass-$M_V$ relation was then varied to give the lowest
residuals with respect to the observed data points.

The comparison between our data and the empirical mass-luminosity relations is
shown in Figure \ref{malu}. Overall, the agreement is quite good and the data
appear to follow the trend of the Henry \& McCarthy (1993) relations. In the
infrared, the points slightly depart from the empirical formulae at brighter
luminosities and masses higher than $\sim 0.85 M_{\odot}$ (depending on the
band). This might be due to evolutionary effects, however we caution that the
Henry \& McCarthy (1993) formulae fit rather noisy data; interestingly, when
the single stars used by Henry \& McCarthy (1993) are overplotted (from their
table 5), at higher masses the agreement is good.  Therefore, considering the
uncertainty in the data and the scatter amongst the empirical relations, the
agreement is within $1 \sigma$ throughout the entire range. Also, the Henry \&
McCarthy (1993) mass-luminosity relation is obtained using stars of
intermediate disk age with various metallicities so that either of these
effects are built in the relations themselves.  We also show the relations of
Delfosse et al. (2000) (valid for $M_K \ge 4.5$) and Kroupa et al. (1993) in
the visible.  In $V$ band the scatter is much larger, but there is no clear
departure from the empirical relations: stars with different metallicities lie
in different parts of the mass-luminosity relation, and less so in the
infrared. This metallicity dependence is also confirmed when the empirical
binary data of Table 3 are overplotted. Such behaviour is predicted by all
theoretical models (e.g. Chabrier \& Baraffe 2000) and it was also noticed by
Delfosse et al. (2000) for stars with masses lower than those covered in the
present study.

In the forth panel of Figure \ref{malu} a linear fit for the stars of Section
\ref{sample} in different metallicity bins is shown. Around and above solar
metallicity (yellow and red lines) the fit is in outstanding agreement with the
binaries of Table 3, whereas a marked difference in the slope appears at low
metallicity (although still in agreement with the empirical relation of Henry
\& McCarthy 1993).  Clearly, more empirical masses and luminosities for metal
poor binaries would be extremely interestingly. If a steeper slope for metal
poor stars is required, that could be achieved by reducing the mass deduced
from the isochrones which in turn would imply a higher helium abundance (see
Figure \ref{hema}). Therefore we suggest that more data on the empirical
mass-luminosity relation for metal poor stars could help to constrain the
helium abundance in stars.

\section{The Helium content in planet host stars}\label{swp}

Stars with planetary companions have been shown to be, on average, considerably
more metal-rich when compared with stars without planets in the solar
neighborhood (e.g. Gonzalez 1997, 1998; Santos, Israelian \& Mayor 2000;
Gonzalez et al. 2001; Santos et al. 2005; Fischer \& Valenti 2005). A high
degree of statistical significance is obtained when iron is used as the
reference element. When other elements are investigated, the situation is much
less clear (see e.g. Gonzalez 2003, 2006 for reviews).  There is some evidence
that planet host stars differ from other nearby stars without planets in their
abundances of Mg, Al, Si, V, Co and Ni. As regards the light elements, no
significant difference between planet and non planet host stars is found for Be
(e.g. Santos et al. 2004b) whereas the situation is more uncertain for Li
(e.g. Gonzalez 2006).

Here we attempt for the first time to derive the helium abundance for a small
set of planetary host stars. They all have metallicities around or above the
solar one, where the isochrones have been proven to be in overall good
agreement with the empirical data (see Section \ref{bitest} and \ref{masslum}).

Our sample of planetary host stars is drawn from the comprehensive list of
Santos et al. (2004a) which provides accurate spectroscopic [Fe/H]
measurements. Abundances for the $\alpha$-elements are available from the same
research group (Gilli et al. 2006). Two planet host stars (HD3651 and HD130322)
were in our original sample of Section \ref{sample}. The other stars have been
chosen if accurate $BV(RI)_C$ colours (from Bessell 1990) were available so
that $T_{eff}$ and bolometric luminosities could be estimated as described in
Section \ref{sample}. If accurate $JHK_S$ magnitudes were also available from
the 2MASS, the IRFM (Casagrande et al. 2006) has been applied directly. We also
used the \emph{Hipparcos} classification to discard variable stars.  The
absence of variability ensures that the stars are likely to be
chromospherically quiet, so that chromospheric-age relations can be more safely
used (Donahue 1998). Most of the planet host stars for which we found accurate
metallicities and photometry have $M_{Bol} < 5.4$, meaning that evolutionary
effects need to be taken into account. For all these stars, age determinations
based on chromospheric indices are available from Saffe, G\'omez \& Chavero
(2005).  Two calibrations are usually adopted to deduce ages from chromospheric
indices: that of Donahue (1993) and that of Rocha-Pinto \& Maciel (1998).  For
a number of reasons, the Donahue (1993) calibration is usually preferred
(Feltzing et al. 2001; Saffe et al. 2005) and is adopted here.

The chromospheric-age relation is considered to be rather robust for ages
younger than $\sim 6$~Gyr (Saffe et al. 2005). When long-term observations are
available --as in Saffe et al. (2005)--, for a given functional form, the age 
uncertainty can be as small as $\sim 1$~Gyr (Donahue 1998). More serious 
problems come if a star is actually
in a Maunder-minimum state where errors estimates can be as high as
5~Gyr. However there are indications that Maunder-minima are very rare among
young stars (e.g. Gustaffson 1999).  In our case, planet host stars have a
median age of 5.1 Gyr using the Donahue (1993) calibration, consistent with the
evidence that most of the nearby solar-type stars have an activity level and
age similar to the Sun (Henry et al. 1996). 
These evolved stars (with typical
$M_{Bol} \sim 4.5$) are more sensitive to the age adopted for the isochrones :
a variation of 1~Gyr changes masses by $0.04-0.05 M_{\odot}$ and helium
abundances $Y$ by $0.02-0.03$ on average, about the same order of the
uncertainties originating from the errors in $T_{eff}$ and $M_{Bol}$.  For
evolved stars we therefore sum in quadrature the errors estimated via
MonteCarlo (according to the prescription given in Section \ref{heres}) to
those resulting from a variation of 1~Gyr in the age of the adopted isochrones.
The complete list of planet host stars with their relevant parameters is shown
in Table \ref{phs}.

\begin{table*}
\setcounter{table}{3}
\centering
\caption{Physical parameters and ages adopted for planet host stars and helium
  abundances and masses deduced from the isochrones.}
\label{phs}
\begin{tabular}{ccccccc}
\hline
HD   & $T_{eff}$ & $M_{Bol}$ & $Z$ & Age (Gyr) & $Y$ & $\frac{M}{M_{\odot}}$\\
\hline
142    & $6247 \pm 81$ & $3.63 \pm 0.05$ & $0.023 \pm 0.003$ & 5.93 & $0.29 \pm 0.04$ & $1.07\pm 0.07$\\
3651   & $5191 \pm 158$& $5.47 \pm 0.07$ & $0.023 \pm 0.003$ & 5.13 & $0.21 \pm 0.06$ & $0.97\pm 0.10$\\
17051  & $6067 \pm 123$& $4.19 \pm 0.05$ & $0.028 \pm 0.003$ & 1.47 & $0.26 \pm 0.07$ & $1.20\pm 0.11$\\
70642  & $5667 \pm 60$ & $4.81 \pm 0.04$ & $0.027 \pm 0.002$ & 3.88 & $0.26 \pm 0.03$ & $1.04\pm 0.07$\\
130322 & $5429 \pm 31$ & $5.54 \pm 0.10$ & $0.017 \pm 0.002$ & 1.24 & $0.25 \pm 0.03$ & $0.86\pm 0.04$\\
160691 & $5689 \pm 46$ & $4.13 \pm 0.04$ & $0.031 \pm 0.002$ & 6.41 & $0.29 \pm 0.03$ & $1.08\pm 0.06$\\
179949 & $6205 \pm 104$& $4.07 \pm 0.05$ & $0.024 \pm 0.002$ & 2.05 & $0.29 \pm 0.06$ & $1.13\pm 0.11$\\
210277 & $5556 \pm 77$ & $4.79 \pm 0.04$ & $0.026 \pm 0.002$ & 6.93 & $0.29 \pm 0.05$ & $0.95\pm 0.07$\\
\hline
\end{tabular}
\begin{minipage}{1\textwidth}
The uncertainty in mass and helium abundance for stars with $M_{bol} < 5.4$ is
not straightforward to determine since age plays a role, as we discuss in the
text.  Errors in $M_{bol}$ also account for the uncertainty in parallaxes.
\end{minipage}
\end{table*}

Our results suggest that within the present day accuracy, planet host stars do
not show any anomalous helium abundance with respect to other field stars in
the same metallicity range. This is in agreement with the fact that $Y_{\odot}$
is recovered within $1 \sigma$ from field stars (Table \ref{ratio}).  A
Kolmogorov-Smirnov test for the helium content between stars with and without
planets in the same metallicity range of Table \ref{phs} confirm that the two
populations are drawn from the same distribution.  The number of available data
points is still small, with large errorbars; but we expect that in the near
future, studies of planet host stars in the lower main sequence will improve on
the robustness of this conclusion.

Most of the studies about exoplanets depend quite strongly on the physical
properties (mostly radius and mass) of the planet host stars. Such properties
are usually obtained from the isochrones. In this study we have proven that at
high metallicity (i.e. around solar) the isochrones can be used with some
confidence. When our masses are compared to those obtained by Santos et
al. (2004a) interpolating isochrones with standard helium abundances, the
agreement is usually good with a mean difference $\Delta M=0.04 \pm 0.13
M_{\odot}$.

\section{Conclusions}\label{conc}

We have compared a set of K dwarfs with accurate effective temperature
$T_{eff}$, metallicity $Z$ and luminosity $M_{bol}$ to a grid of stellar
theoretical isochrones, in order to indirectly determine the helium abundance
$Y$ in the stars. We then derive the helium-to-metal enrichment ratio $\Delta Y
/ \Delta Z$ in the Solar Neighbourhood.

For all our stars the fundamental physical parameters were derived empirically
and homogeneously, with the specific aim of measuring small differential
effects along the lower main sequence (Section \ref{sample}).  The isochrones
used are among the most up-to-date, implementing the latest input physics and
covering a large grid of $(Y,Z)$ values; also, we have verified that, for
similar $(Y,Z)$ content they compare very well to other recent sets
(Yonsei-Yale, Teramo and MacDonald) and therefore our results do not depend
significantly on the specific isochrones employed (Section \ref{Padova}).

While we derive reasonable values for helium content in our K dwarfs around
solar metallicity, the isochrones yield very low values of helium in the metal
poor K dwarfs. At metallicities around and above the solar, we obtain $\Delta Y
/ \Delta Z \sim 2$ (Section \ref{heres}), but at low metallicities, the match
between the theoretical and observed main sequence is so poor.  Theoretical
isochrones can be forced to fit metal poor stars only by assuming {\it very low
helium abundances}, well below the primordial Big-Bang value (Section
\ref{method}; Figure \ref{yz}). This result is quite puzzling.

Although different isochrone sets exhibit some differences in the derived
helium content (especially at the lowest metallicities and luminosities), very
low helium abundances are already found for moderately metal poor and bright
stars, where all the isochrones agree remarkably well.  The size and
homogeneity of the sample of stars, the accuracy in both empirical and
theoretical data, together with the possibility of making the comparison
directly in the $T_{eff} - M_{Bol}$ plane, where the effect of the helium
content are most evident, are the major improvements over similar work in the
past.

The helium discrepancy was known to exist (Lebreton et al. 1999) -- the
present sample greatly extends on previous work and appears to show that there
is a clear problem in fitting stellar models to low mass, metal poor stars.
Lebreton et al. (1999) showed that adjustments in stellar models (diffusion)
and observations (NLTE effects) alleviate the problem but do not solve it,
although their conclusions were based on a rather small number of stars. 

We also find that diffusion and NLTE are unlikely to solve the problem
completely. We rule out systematic errors in our metallicity scale or
temperature scale as the culprit, since the systematic errors required would be
much larger than our error estimates comfortably allow, and external checks
indicate our other derived stellar physical parameters (mass, luminosity and
temperature) are excellent.  Our very low helium abundances in the metal poor K
dwarfs can be avoided via the ad hoc assumption that the mixing-length
parameter decreases with decreasing metallicity for $Z$ below solar;
this is of course a very major change to make to stellar models.

Interestingly, discrepancies between theory and observations for stars less
massive than the Sun have already been reported in the literature from the
studies of binaries (e.g. Popper 1997; Torres \& Ribas 2002, see Section
\ref{bitest}).  However, such discrepancies in the range of K dwarfs have not
yet caught --in our opinion-- the attention they deserve, as most of the
studies aiming to empirically measure mass-luminosity relations have
preferentially focused on either earlier or later spectral types.  We urge
stellar model makers to reassess the modeling of low metallicity, lower main
sequence stars. 

As we have discussed in Section \ref{bitest}, accurate masses and fundamental
physical parameters for metal poor dwarf binaries would be a powerful test of
stellar models. The slope of the mass-luminosity relation in the metal poor
regime could directly test the effect of diffusion (Section \ref{notguilty})
and be used to determine the helium abundance (Section \ref{masslum}). Low
degree modes from space-based asteroseismology missions can be used to
determine the helium abundance in stellar envelopes with an accuracy of 0.03
for a $0.8 M_{\odot}$ star (Basu et al. 2004). Such accuracy is comparable to
that in the present work. The study of such modes in metal poor dwarfs with
on-going or forthcoming space missions like \emph{COROT} and \emph{Kepler} is
therefore urged.

Theoretical isochrones are extensively used to determine the distance scale
fitting the observed colour-magnitude diagram of globular clusters.  Metal poor
isochrones with primordial, or close to primordial, helium abundance are used
to infer their distance.  However, our direct observations of nearby stars are
challenging low metallicity isochrones : if extremely helium poor isochrones
are formally needed to fit empirical data, this would stretch the
(isochrone-based) distance scale by $0.1-0.2$ magnitudes (but note that
empirically calibrated distance scales based on nearby subdwarfs would remain
unaffected). Besides, assuming primordial helium for the metal poor population
in such objects would imply, in a differential sense, extremely helium
enhanced values for the metal rich counterparts. 
We stress that here we are not arguing that our very low helium
abundances are real; rather we attribute them to the current limits in stellar
models. Clearly direct parallaxes measurements for stars in globular clusters
(as \emph{Gaia} and \emph{SIM} will provide) will shed new light on the
problem. 

In the meanwhile, more studies for modeling low mass, metal poor stars are
needed. At present, the impact of 3D model atmospheres on stellar abundance
determinations is revolutionizing the field; the solar model itself is under
profound revision, and in coming years we expect many exciting breakthroughs.

Our study shows that at metallicities around and above solar, theoretical
models are in good agreement with observations. This is of interest to studies
of exoplanets, which are primarily around host stars of about the solar
metallicity, since they still heavily depend on theoretical stellar models to
constrain properties of the parent star.  

\section*{Acknowledgments}

LC acknowledges the hospitality of the Department of Astronomy at the
University of Pennsylvania, the Research School of Astronomy and Astrophysics
at Mount Stromlo, the University of Padova and the Osservatorio Astronomico di
Padova, where part of this work was carried out.  LC acknowledges the Finnish
Graduate School in Astronomy and Space Physics, the Turku University Foundation
and the Otto A. Malm Foundation for financial support.  This study was funded
also by the Academy of Finland (CF) and by the $6^{\textrm{\tiny{th}}}$
Framework Program of the European Communities (Marie Curie Fellowship
nr. MEIF-CT-2005-010884; LP).  We thank Regner Trampedach for very useful and
constructive discussions on the mixing-length theory.  We are also indebted to
Yveline Lebreton and the referee Jo\~ao Fernandes for a very careful reading 
of the
manuscript and many constructive comments that significantly improved the 
paper.  The
research has made use of the the General Catalogue of Photometric Data operated
at the University of Lausanne and the SIMBAD database, operated at CDS,
Strasbourg, France. The publication makes use of data products from the Two
Micron All Sky Survey, which is a joint project of the University of
Massachusetts and the Infrared Processing and Analysis Center/California
Institute of Technology, funded by the National Aeronautics and Space
Administration and the National Science Foundation.

\appendix

\section[]{The metal mass fraction}

The metal mass fraction $Z$ is related to the measured abundance [M/H] by 
\begin{equation}
\textrm{[M/H]}=\log(Z/X)-\log(Z/X)_{\odot}
\end{equation}
where $Z=1-X-Y$ and $X$ and $Y$ are the hydrogen and helium mass fraction,
respectively. It follows that for any given star with measured [M/H] the
corresponding metal mass fraction is
\begin{equation}\label{correction}
Z = \epsilon \, Z_{\odot} 10^{\textrm{[M/H]}}
\end{equation}
where 
\begin{equation}
\epsilon = \frac{(1-Y)/X_{\odot}}{1+(Z/X)_{\odot}10^{\textrm{[M/H]}}}. 
\end{equation} 

\begin{figure}
\begin{center}
\includegraphics[scale=0.47]{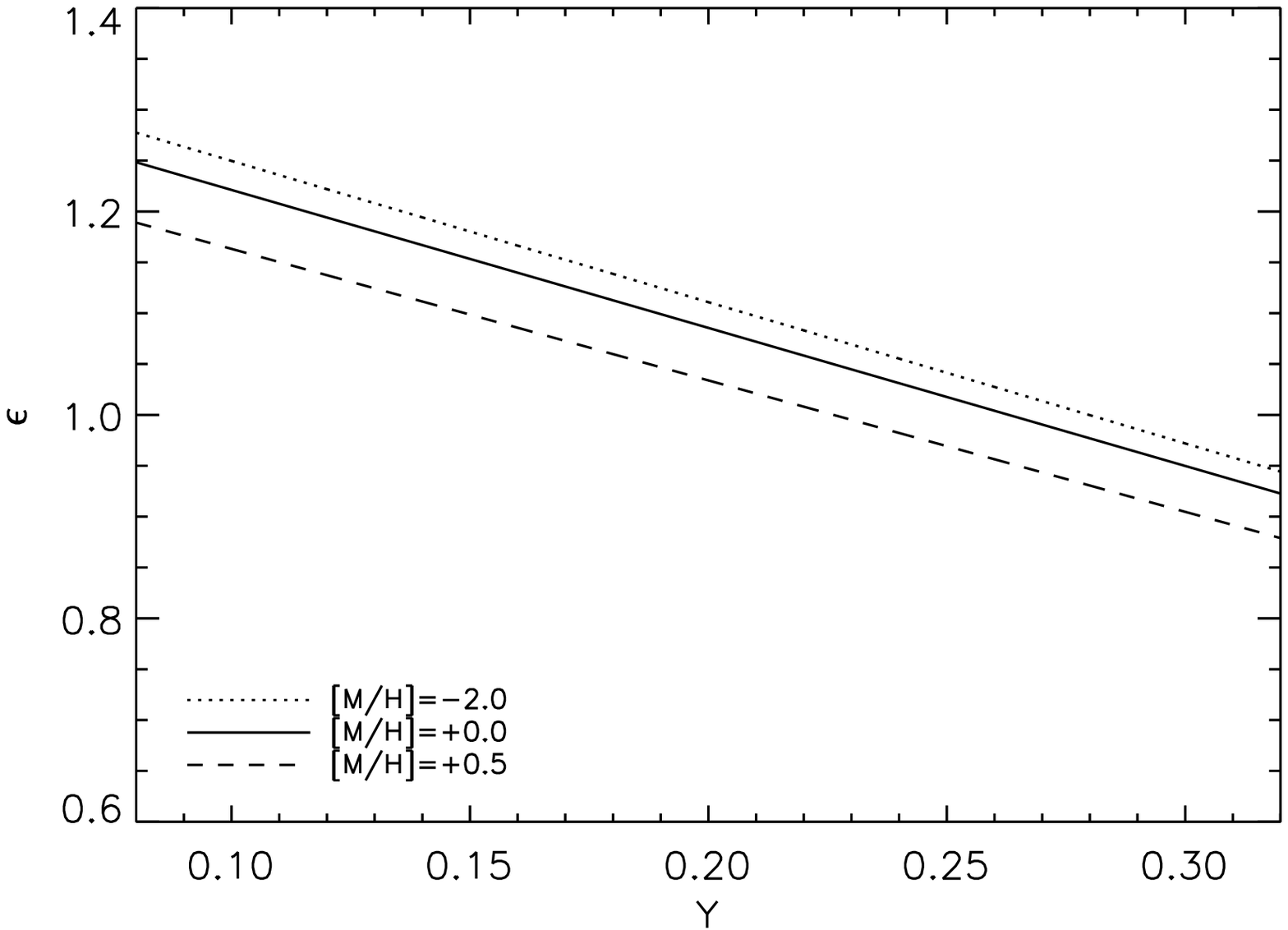}
\caption{{Change in metal mass fraction for different [M/H] and helium
    abundances $Y$.}}
\label{ApFi}
\end{center}
\end{figure}

From Figure (\ref{ApFi}) it is clear that the use of such a correction has
negligible effect (few percent) for standard helium values and up to $\sim 25$
percent in the case of an helium abundance as low as $Y=0.10$.  For all the
stars in our sample, we have computed $Z$ and $Y$ iteratively: $Z=Z_{\odot}
\times 10^{\textrm{[M/H]}}$ was used to obtain a first estimate of the metal
mass fraction used for interpolating over our grid of isochrones as explained
in Section \ref{method}.  This returned an estimate of $Y$ that was then used
into equation (\ref{correction}) and the newly computed metal mass fraction
used for another interpolation over our grid of isochrones.  The procedure was
iterated until $Z$ and $Y$ converged to better than 0.001, usually within $4-5$
iterations.

We have adopted the solar abundances of Section \ref{Padova}, but notice that
changing from $(Z/X)_{\odot}=0.0236$ to $(Z/X)_{\odot}=0.0245$ affects the
final values of $Z$ and $Y$ by less than $10^{-4}$.

\section[]{The Lutz-Kelker bias}

The Lutz--Kelker bias on the absolute magnitude is in principle present at any
level of parallax accuracy.  The bias has two components, the first of which is
statistical: since the number of stars increases with decreasing parallaxes
(i.e. larger distances and sampled volumes) observational errors on the
parallax will not cancel out exactly, giving a net effect of more stars with
overestimated parallaxes. This makes the correction on the magnitudes of the
individual stars statistical and dependent upon the properties of the sample.
The second component of the bias arises because distances (and hence absolute
magnitudes) are not linear functions of the parallaxes. Even if errors have a
normal distribution in parallax, they propagate to a skewed distribution in
distance (absolute magnitude). Once again the effect is to favor more distant
(brighter) stars to appear closer (fainter).  It is clear that this is a
correction that has to be applied to the absolute magnitude of individual
stars.

In recent literature there is a certain degree of confusion about what exactly
is the Lutz--Kelker bias, whether any correction should be applied and, if so,
how large it should be. The value of the corrections to be applied depends on
the distribution of the true parallaxes $\omega_{0}$. In principle the true
parallax distribution could be derived by deconvolving the observed
distribution for observational errors. In practice, the deconvolution process
is quite uncertain and most of the authors prefer to use analytical formulae
for the correction. The most widely used is that of Hanson (1979), who gave
analytical formulae relating absolute magnitude corrections to the proper
motion distribution of the sample of stars. However, the use of such
formulation is highly risky and it does not provide the necessary accuracy
because of the strong dependence on the variety of input parameters. As
recommended by Brown et al. (1997) in their paper on the properties of the
\emph{Hipparcos} catalog, any correction should be tailored for each specific
case. In fact, the blind use of Hanson's formulation is often counterproductive
as noticed already by many authors (Gratton et al. 1997; Sandage \& Saha
2002). Assuming a uniform density case, Hanson's formulation can be used as an
estimator for the worst-case scenario, but not as a correction for the bias.
Since the \emph{Hipparcos} catalog completeness decreases with increasing
magnitude, the statistical correction is negligible.

On the other hand, since we are dealing with the magnitude difference between
individual stars and a reference isochrone, the correction for the skewed
distribution is more relevant (at least in principle). Since magnitudes (and
distances) are not linear functions of the parallax, one needs to resort to
likelihood methods where the complete probability distribution function (pdf)
of $\omega$ given its uncertainties, is transformed into the corresponding
distribution function for distance and magnitude. By means of such a method we
prove the bias to be well within our observational errors.

Assuming a Gaussian distribution for the errors around the measured parallax
$\omega$ 
\begin{equation}
f\left(\tilde{\omega}\right)=\frac{1}{\sqrt{2\pi}\sigma}e^{-\frac{(\tilde{\omega}-\omega)^2}{2\sigma^2}},
\end{equation}
the expectation value for the distance $R$ is 
\begin{equation}\label{asymp}
E[R|\omega]=
\int_{-\infty}^{+\infty}\frac{1}{\tilde{\omega}}f\left(\tilde{\omega}\right) \textrm{d}\tilde{\omega}=
\frac{1}{\sqrt{2\pi}\sigma}\int_{-\infty}^{+\infty}g(\tilde{\omega})e^{\frac{t(\tilde{\omega})}{{2\sigma^2}}}\textrm{d}\tilde{\omega}
\end{equation}
where $g(\tilde{\omega})=1/\tilde{\omega}$ and $t(\tilde{\omega})=-(\tilde{\omega}-\omega)^{2}$. Defining 
$u=\tilde{\omega}-\omega$ and expanding $g(\tilde{\omega})$ around the maxima 
of $t(\tilde{\omega})$ one gets the following series
\begin{equation}
g(\tilde{\omega})=\frac{1}{\omega}\sum_{n=0}^{\infty}(-1)^{n}\left(\frac{u}{\omega}\right)^n.
\end{equation}
Given the accuracy of our parallaxes, the condition $|u/\omega| < 1$ for the
expansion is certainly satisfied within, say, $3\sigma$.  The choice of the
$3\sigma$ cutoff sounds perfectly reasonable for the purposes of our
calculation.  Alternatively, a more rigorous approach has been investigated by
Smith \& Eichhorn (1996).

Once the series expansion for $g(\tilde{\omega})$ is known, the integral in 
eq. (\ref{asymp}) reads  
\begin{equation}
E[R|\omega]=\frac{1}{\sqrt{2\pi}\sigma}\int_{-\infty}^{+\infty}
\frac{1}{\omega}\sum_{n=0}^{\infty}(-1)^{n}\left(\frac{u}{\omega}\right)^n
e^{-\frac{u^{2}}{2\sigma^{2}}} \textrm{d}u
\end{equation}
and can easily be calculated by using Gaussian integrals and noticing that odd
terms vanish. The result is the following series
\begin{equation}
E[R|\omega]=\frac{1}{\omega}\sum_{n=0}^{\infty}\frac{(2n)!}{2^{n}n!}\left(\frac{\sigma}{\omega}\right)^{2n}
\end{equation}
and the amplitude of the bias 
\begin{equation}
E[R|\omega]-\frac{1}{\omega}\sim\frac{1}{\omega}\left(\frac{\sigma}
{\omega}\right)^{2} 
\end{equation}
can be easily checked to be negligible for our adopted accuracy in parallaxes
(6\%).  

In the same manner we can give an estimate for the bias in absolute
magnitudes $M_{\xi}$ in a given band $\xi$ or bolometric. We thus have
\begin{equation}
E[M_{\xi}|\omega]=\frac{1}{\sqrt{2\pi}\sigma}\int_{-\infty}^{+\infty}g(\tilde{\omega})e^{\frac{t(\tilde{\omega})}{{2\sigma^2}}}\textrm{d}\tilde{\omega},
\end{equation}
where now, for a given apparent magnitude $\xi$, we have
\begin{equation}
g(\tilde{\omega})=\xi+5\log(\tilde{\omega})-10.
\end{equation}
Hence the integral to be computed is 
\begin{displaymath}
E[M_{\xi}|\omega]=\frac{1}{\sqrt{2\pi}\sigma}\int_{-\infty}^{+\infty}\Big[\xi+5\log(\omega)-10+
\end{displaymath}
\begin{equation}
\phantom{E[M_{\xi}|\omega]=}5\log(e)\sum_{n=1}^{\infty}\frac{(-1)^{n+1}}{n}\left(\frac{u}{\omega}\right)^{n}\Big]e^{-\frac{u^{2}}{2\sigma^{2}}}\textrm{d}u
\end{equation}
and gives the following bias
\begin{equation}
E[M_{\xi}|\omega]-M_{\xi}=-5\log(e)\sum_{n=1}^{\infty}\frac{(2n-1)!}{2^{n}n!}\left(\frac{\sigma}{\omega}\right)^{2n}.
\end{equation}
Also for the absolute magnitudes the bias is negligible. At the first order
the above equation gives 
\begin{equation}
E[M_{\xi}|\omega]-M_{\xi}=-\frac{5}{2}\log(e)\left(\frac{\sigma}{\omega}\right)^{2}
\sim -0.004\; \textrm{mag}
\end{equation}
in the case of lowest parallax accuracy and thus well within our observational 
errors.

We finally note that given the high precision of our parallaxes ($\sigma/\omega
\le 0.06$) a fully Bayesian approach is not needed, also considering that it
would require a priori assumptions on the parameter distributions. A Bayesian
approach is indeed demanded for lower precision parallaxes and it has been
extensively studied by Smith (1985, 1987, 2003) in his series of papers on the
Lutz--Kelker bias.

\end{document}